\documentclass[11pt]{article}

\usepackage{graphicx,cite}
\usepackage{amsmath}
\usepackage{amsfonts}
\usepackage{amssymb}
\usepackage{rotating}
\usepackage{booktabs}
\newcommand{\ra}[1]{\renewcommand{\arraystretch}{#1}}
\usepackage{xcolor}
\usepackage{soul}

\newcommand{\eV}{\ensuremath{\mathrm{eV}}}
\newcommand{\MeV}{\ensuremath{\mathrm{MeV}}}
\newcommand{\GeV}{\ensuremath{\mathrm{GeV}}}
\newcommand{\tr}{\operatorname{tr}}
\newcommand{\diag}{\operatorname{diag}}
\newcommand{\SU}{\ensuremath{\mathrm{SU}}}
\newcommand{\U}{\ensuremath{\mathrm{U}}}

\renewcommand{\O}{\ensuremath{\mathrm{O}}}
\renewcommand{\Im}{\ensuremath{\operatorname{Im}}}

\makeatletter
\DeclareRobustCommand*{\bfseries}{%
  \not@math@alphabet\bfseries\mathbf
  \fontseries\bfdefault\selectfont
  \boldmath
}
\makeatother

\begin{document}

\begin{titlepage}

\vspace*{-15mm}
\begin{flushright}
TTP14-029
\end{flushright}
\vspace*{0.7cm}

\begin{center}
{
\bfseries\LARGE
The double mass hierarchy pattern: \\
simultaneously understanding \\
quark and lepton mixing
}
\\[8mm]
Wolfgang Gregor Hollik$^*$
\footnote{E-mail: \texttt{wolfgang.hollik@kit.edu}}
and
Ulises Jes\'us Salda\~na Salazar$^{*,\dag}$
\footnote{E-mail: \texttt{ulisesjesus@fisica.unam.mx}}
\\[1mm]
\end{center}
\vspace*{0.30cm}
\centerline{$^*$ \itshape
Institut f\"ur Theoretische Teilchenphysik, Karlsruhe Institute of Technology}
\centerline{\itshape
Engesserstra\ss{}e 7, D-76131 Karlsruhe, Germany}
\vspace*{0.50cm}
\centerline{$^\dag$ \itshape
Instituto de F\'{\i}sica, Universidad Nacional Aut\'onoma de M\'exico}
\centerline{\itshape
Apdo. Postal 20-364, 01000, M\'exico D.F., M\'exico}
\vspace*{0.80cm}

\begin{abstract}
\noindent
The charged fermion masses of the three generations exhibit the two
strong hierarchies $m_{3} \gg m_2 \gg m_1$. We assume that also neutrino
masses satisfy \(m_{\nu 3} > m_{\nu 2} > m_{\nu 1}\) and derive the
consequences of the hierarchical spectra on the fermionic mixing
patterns. The quark and lepton mixing matrices are built in a general
framework with their matrix elements expressed in terms of the four
fermion mass ratios $m_u/m_c$, $m_c/m_t$, $m_d/m_s$, and $m_s/m_b$ and
$m_e/m_\mu$, $m_\mu/m_\tau$, $m_{\nu 1}/m_{\nu 2}$, and $m_{\nu
  2}/m_{\nu 3}$, for the quark and lepton sector, respectively. In this
framework, we show that the resulting mixing matrices are consistent
with data for both quarks and leptons, despite the large leptonic mixing
angles. The minimal assumption we take is the one of hierarchical masses
and minimal flavour symmetry breaking that strongly follows from
phenomenology. No special structure of the mass matrices has to be
assumed that cannot be motivated by this minimal assumption. This
analysis allows us to predict the neutrino mass spectrum and set the
mass of the lightest neutrino well below \(0.01\,\eV\). The method also
gives the $1\,\sigma$ allowed ranges for the leptonic mixing matrix
elements. Contrary to the common expectation, leptonic mixing angles are
found to be determined solely by the four leptonic mass ratios without
any relation to symmetry considerations as commonly used in flavor model
building. Still, our formulae can be used to build up a flavor model
that predicts the observed hierarchies in the masses---the mixing
follows then from the procedure which is developed in this work.
\end{abstract}

\end{titlepage}

\setcounter{footnote}{0}

\tableofcontents
\newpage

\section{Introduction}
The Standard Model of particle physics (SM) describes the interactions
among elementary particles at high energies with great success. In spite
of this, the setup of the SM lacks an explanation of the origin of
fermion masses and mixing. In particular, for the quark sector, one
observes six masses, three mixing angles and one phase. It is a simple
exercise to relate the quark mixing matrix to the fundamental parameters
of the theory, the Yukawa couplings. Generally, however, it is said that
mixing angles as well as the masses are \emph{independent} free
parameters. Is there really no functional relation between the quark
masses and the corresponding mixing matrix elements? There are many
models in the literature that want to give an explanation of the mixing
matrix elements in terms of the masses \cite{Gatto:1968ss,
  Cabibbo:1968vn, Tanaka:1969bw, Oakes:1969vm, Genz:1973sn,
  Pagels:1974qg, Ebrahim:1977hb, Fritzsch:1977za, Weinberg:1977hb,
  DeRujula:1977ry, Wilczek:1977uh, Mohapatra:1977rj, Wilczek:1978xi,
  Wyler:1978fj, Fritzsch:1986sn, Frampton:1991aya, Rosner:1992qa,
  Raby:1995uv, Ito:1996zt, Xing:1996hi, Xue:1996fm, Barbieri:1996ww,
  Falcone:1998us, Mondragon:1998gy, Mondragon:1999jt, Fritzsch:1999ee,
  Branco:2010tx, Canales:2013cga}.  Most of them put assumptions on a
specific texture in the original mass matrices. We shall show, by
contrast, that the pure phenomenological observation of strong
hierarchies in the quark masses leads to a functional description of the
mixing matrix elements in terms of mass ratios. The consequences in the
mixing of this phenomenological observation have already been
studied~\cite{Fritzsch:1986sn, Hall:1993ni, Xing:1996hi, Rasin:1997pn,
  Rasin:1998je, Fritzsch:1999rb, Fritzsch:1999ee, Xing:2012zv}. Our
approach differs from the previous ones in many aspects: i) we take the
Singular Value Decomposition of the complex mass matrices as a starting
point offering a generic treatment for both quarks and leptons; ii) by
means of an approximation theorem we mathematically formulate the steps
to build the reparametrization of the mixing matrix in terms of the
singular values (fermion masses); iii) we rotate the mass matrices in
all three planes of family space whereas before, the 1-3 rotation was
neglected; iv) as the two unitary rotations in the 2-3 and 1-3 plane
involve an approximation ($m_{f,1} =0$ and $m_{f,2}= 0$, respectively)
we consider for the first time a modified method of perturbation theory
to add the effect of the terms neglected; v) we do not consider the
complex CP phases as free parameters and show that a minimal choice is
sufficient to explain CP data; vi) we provide explicit formulae for the
mixing angles in terms of only mass ratios.

The applicability of this formulation to the leptonic mixing is not
clear \emph{a priori}. First, neutrino masses do not show any strong
hierarchy, at best a very mild one. Second, the leptonic mixing matrix
exhibits large mixing, while the one in the quark sector is rather close
to the unit matrix. This picture seems to suggest two quite different
origins for the respective mixing matrices: quark masses strongly
dominating the mixing patterns, whereas geometrical factors found from
symmetries shaping the leptonic mixing, with only a weak intervention
from the lepton masses~\cite{Ishimori:2010au, King:2013eh}.

Fermion masses, on the other hand, are also as puzzling as the mixing
matrices: the top quark mass is by far the largest among the charged
fermions, there are six orders of magnitude separating the top quark
from the electron mass, six orders of magnitude separating the largest
neutrino mass from the electron mass (assuming a neutrino mass scale of
\(0.1\,\eV\)). There are three orders of magnitude between the masses of
the up-type quarks, whereas two orders of magnitudes in the down-quark
sector. Top and bottom quark are separated by two orders of
magnitude---the lightest charged lepton and the heaviest quark by again
six orders of magnitude. Within each (charged) fermion species ($f = u,
d, e$), the masses follow a hierarchy $m_{f,3} \gg m_{f,2} \gg m_{f,1}$,
\begin{equation}
  \begin{aligned}
    m_u : m_c : m_t \approx 10^{-6} : 10^{-3} : 1, &\quad\quad
    m_d : m_s : m_b \approx 10^{-4} : 10^{-2} : 1,  \\
    m_e : m_{\mu} : m_{\tau} &\approx 10^{-4} : 10^{-2} : 1,
  \end{aligned}
\end{equation}
while the two squared mass differences measured from neutrino
oscillations obey a much weaker hierarchy,
\begin{eqnarray}
  \Delta m_{21}^2 : \Delta m_{31(32)}^2 \approx 10^{-2} : 1.
\end{eqnarray}

Quark masses plus mixing parameters sum up to ten arbitrary physical
parameters in the SM.  Consideration of neutrino masses, whether Dirac
or Majorana, adds at least ten more parameters to the count. Two more
complex phases and a possibly arbitrary number of masses for sterile
neutrinos appear in the more general cases including Majorana
neutrinos~\cite{Schechter:1980gr}. The SM \textit{per se} seems to
lack a course of action on how to relate the mixing matrix elements to
the corresponding fermion masses.

The first realization of a mixing angle in terms of the masses is
commonly assigned to Gatto et al.~\cite{Gatto:1968ss} which is referred
to as the Gatto-Sartori-Tonin relation.  This relation is an
expression of the Cabibbo angle commonly written as
\begin{equation}\label{eq:GSTrel}
  \theta_{12}^q \approx \sqrt{\frac{m_d}{m_s}},
\end{equation}
where originally, the authors of~\cite{Gatto:1968ss} found a similar
relation in terms of light meson masses from the demand of weak
self-masses being free from quadratic divergences. In a footnote, they
break it down to an elementary discussion in a ``naive quark model'' and
state
\begin{equation}
  \tan^2\theta = \frac{m_n - m_p}{m_p} = \frac{m_n}{m_\lambda},
\end{equation}
where \(m_n\), \(m_p\), and $m_\lambda$ are the old notations of down-,
up-, and strange-quark masses (moreover, the second equal sign was
misleadingly written as a minus sign). The first work referring to
\cite{Gatto:1968ss} as origin of ``\(\tan\theta = m_n / m_\lambda \)''
was \cite{Tanaka:1969bw} (even though with a typo in the abstract). For
small angles, \(\tan\theta \approx \theta\) and we are at
Eq.~\eqref{eq:GSTrel}. Since \(\sqrt{m_d / m_s}\) is an astonishingly
good approximation for the Cabibbo angle, we will show in the course of
this paper how to rearrive at this expression in a formal way of
parametrizing mixing matrices in terms of invariants.

The work of \cite{Gatto:1968ss} was followed by derivations of the same
formula focused on the derivation in a more model-building related
approach using left-right symmetric scenarios~\cite{Gatto:1968ss,
  Cabibbo:1968vn, Tanaka:1969bw, Oakes:1969vm, Genz:1973sn,
  Mohapatra:1977rj, Fritzsch:1979zq}. In the same decade, a model
independent approach was initiated where mass matrices with different
null matrix elements (``texture zeros'') were
considered~\cite{Fritzsch:1977vd, Ramond:1993kv, Branco:1999nb,
  Roberts:2001zy, Fritzsch:2002ga, Gupta:2012dma}; similar relations
were then found for other mixing angles. Subsequently, horizontal or
family discrete symmetries were used in order to relate the three
families in a non-trivial fashion~\cite{Pakvasa:1975ti, Pakvasa:1977in,
  Derman:1978rx, Wyler:1978fj, Yamanaka:1981pa, Yahalom:1983kf,
  Wilczek:1977uh, Wilczek:1978xi}. In their initial stage, though, the
experimental uncertainty in the mixing angles and fermion masses was
still too large as to build a stable model consistent with the unstable
phenomenology. This approach was vigorously resurrected in the last
decade when precision measurements for neutrino oscillations started
\cite{King:2013eh, Altarelli:2014dca, King:2014nza}. Relations between
the neutrino mixing angles and lepton mass hierarchies were
found~\cite{Fritzsch:2006sm, Fritzsch:2009sm} where the values for the
three neutrino masses are compatible with what follows from our method,
though \(\theta_{13}\) was predicted too low (only about \(3^\circ\)).
Nevertheless, up to now, no complex mass matrix with a well-motivated
constrained set of parameters has been found to entirely and
successfully postdict the Cabibbo-Kobayashi-Maskawa (CKM) quark mixing
matrix or to predict the Pontecorvo-Maki-Nakagawa-Sakata (PMNS) matrix
in the lepton sector. In this work, we do not focus on a specific model
predicting mixing angles, but give explicit relations following from a
model independent treatment based on the observation of the two strong
hierarchies \(m_3 \gg m_2 \gg m_1\) in the charged fermion
masses. Moreover, we dare to apply the same fomulae to the neutrino
mixing and derive the PMNS angles with astonishingly good agreement.

This paper is organized in the following way: first, we start discussing
the generic treatment of mixing matrices following from hierarchical
mass matrices in Section~\ref{sec:massmix}, where we focus on the
mathematical derivation of relations among fermion mass ratios and
mixing angles. This result gets applied to the phenomenological data in
Section~\ref{sec:applpheno}. Finally, we conclude. In the appendices, we
review the current status of input data, give a brief statement about
the applicability of the method elaborated in this work, comment on the
hierarchical structure of the mass matrices as a consequence of
hierarchical masses and minimal flavor symmetry breaking, and provide
the explicit, approximative formulae that gave the results of
Section~\ref{sec:applpheno}.

\section{Mass and mixing matrices}
\label{sec:massmix}
Let us extend the SM by three right-handed neutrinos to have a more
symmetric treatment of the problem in the quark and lepton sector. Dirac
neutrinos alone still leave the question open why the Yukawa couplings
for neutrinos are so much smaller than for the charged
fermions. Nonetheless, in the description of fermion mixings in terms of
fermion masses this assumption does not play a r\^ole and later we take
an effective neutrino mass matrix without the need to specify whether
neutrinos are Dirac or Majorana. The most general, renormalizable and
gauge invariant construction of fermion mass matrices follows from the
Yukawa Lagrangian
\begin{eqnarray}
  -{\cal L}_Y = \sum_{f=d,e}
  {\cal Y}_f^{ij} \, \overline{\psi}_{fL,i} \, \Phi \, \psi_{fR,j} +
  \sum_{f=u,\nu} {\cal Y}_f^{ij} \,
  \overline{\psi}_{fL,i} \, (i\sigma_2\Phi^*) \, \psi_{fR,j} + {\text{H.c.}},
\end{eqnarray}
where $i,j = 1,2,3$ are family indices and summation over them is
implicitly understood. The generic fermion fields are denoted as
\(\psi_f\), where the left-handed fermions are grouped into \(\SU(2)_L\)
doublets and the right-handed ones are the usual singlets. The Higgs
doublet is given by \(\Phi = (\phi^+, \phi^0)\) whereas its nonvanishing
vacuum expectation value \(v = \langle \phi^0 \rangle = 174\,\GeV\). The
spontaneous breakdown of electroweak symmetry gives rise to four Dirac
mass matrices of the form
\begin{eqnarray}
  {\cal M}_f = v {\cal Y}_f.
\end{eqnarray}
These mass matrices are $3 \times 3$ complex arbitrary matrices; each of
them is diagonalized by a biunitary transformation
\begin{eqnarray}\label{eq:SVD}
  D_f = L^f {\cal M}_f {R^f}^\dagger,
\end{eqnarray}
where $D_f$ is a diagonal matrix with real and positive entries while
$L^f$ and $R^f$ are two unitary matrices acting in family space on left-
and right-handed fermions of type \(f\) respectively. Both
transformations, $L^f$ and $R^f$, correspond to the unitary matrices
appearing in the \emph{Singular Value Decomposition} of ${\cal
  M}_f$. These unitary matrices transform the sets of three left- or
three right-handed fermion fields each from the interaction basis to the
physical mass basis
\begin{eqnarray}
  \psi^\prime_{f,L} = L^f \psi_{f,L} \quad\quad {\text{and}} \quad\quad
  \psi^\prime_{f,R} = R^f \psi_{f,R}.
\end{eqnarray}
The mass eigenstates are therewith \(\psi^\prime_f\). In return, the
diagonal weak charged current interactions are no longer diagonal, and
mix different fermion families. This occurs as a consequence of the
mismatch between the two different left unitary matrices acting inside
the same fermion sector which results in the observable mixing matrices
in the charged current interactions
\begin{eqnarray}
  V_{\text{CKM}} = L^{u} {L^{d}}^\dagger \quad\quad
  {\text{and}} \quad\quad
  U_\text{PMNS} = L^e {L^{\nu}}^\dagger.
\end{eqnarray}

\subsection{The double mass hierarchy pattern (DMHP)}
The singular values of the diagonal matrix \(D_f\) in Eq.~\eqref{eq:SVD}
are to be identified with the measured fermion masses (see
\ref{app:data}). An interesting and not yet exploited fact is that
the observed hierarchies in the masses (singular values) can be used to
approximate the original mass matrices by lower-rank matrices as stated
in the Schmidt-Mirsky approximation
theorem~\cite{Schmidt1907433,EckartYoung,Mirsky,Golub1987317}.\footnote{It
  is often wrongly called the Eckart-Young-Mirsky or simply Eckart-Young
  theorem, see~\cite{Stewart1993551} for an early history on the
  Singular Value Decomposition.}

The left and right unitary matrices, $L^f$ and $R^f$ are decomposed into
the left and right singular vectors, $l_{f,i}$ and $r_{f,i}$
($i=1,2,3$), and built up as ${L^f}^\dagger = [l_{f,1}, l_{f,2},
l_{f,3}]$ and ${R^f}^\dagger = [r_{f,1}, r_{f,2}, r_{f,3}]$. Each pair
of singular vectors correspond to the singular value $m_{f,i}$. For
square matrices when all three singular values can be ordered as
$m_{f,3} >m_{f,2} > m_{f,1} \geq 0$, the decomposition is unique up to a
shared complex phase for each pair of singular vectors.\footnote{In the
  case of degeneracy among some of the singular values, there is no
  longer a unique Singular Value Decomposition for ${\cal M}_f$.  This
  matters in the discussion of degenerate neutrino masses.}

The number of non-zero singular values equals the rank of the mass
matrix ${\cal M}_f$. The mass matrix can be written in terms of its
singular values with the respective left and right singular vectors as a
sum of rank one matrices,
\begin{eqnarray} \label{eq:dmhp}
{\cal M}_f = \left[ \left( l_{f,1}
      \frac{m_{f,1}}{m_{f,2}} r_{f,1}^\dagger + l_{f,2}
      r_{f,2}^\dagger\right)\frac{m_{f,2}}{m_{f,3}} + l_{f,3}
    r_{f,3}^\dagger\right]m_{f,3}.
\end{eqnarray}
Any hierarchy among the singular values is of major interest to us as it
leads to a lower-rank approximation ${\cal M}_f^{r}$
($r={\text{rank}}[{\cal M}_f^{r}] <3$). The lower-rank approximation is
the closest matrix of the given rank to the original matrix, where
``close'' has to be specified (see \ref{app:applic}). We obtain it
by keeping the largest singular values and setting the smaller ones
equal to zero. The lower rank matrices are unique if and only if all
the kept singular values are larger than those set to zero.

Because of $m_{f,3} \gg m_{f,2} \gg m_{f,1}$, Eq.~\ref{eq:dmhp} provides
a powerful way to appreciate the double hierarchy of its singular values
and the emerging relation to its rank by the use of Schmidt-Mirsky's
approximation theorem.  As both types of quarks and charged lepton
masses satisfy those two hierarchies, we conclude, that their mass
matrices can be safely approximated as either matrices of rank one or
rank two, depending on how strong their double mass hierarchy pattern
(DMHP) is.

As illustrated in Eq.~\eqref{eq:dmhp}, this expression points also to
the fact that the fermion mass ratios $m_{f,1}/m_{f,2}$ and
$m_{f,2}/m_{f,3}$ play the dominant r\^ole in determining the structure of
the mass matrix whereas $m_{f,3}$ sets the overall mass scale. Only
those two ratios will be necessary in the determination of the mixing
parameters, since the overall mass scale can be factored out. For later
use, we abbreviate $\hat{m}_{f,1} = m_{f,1}/m_{f,3}$ and $\hat{m}_{f,2}
= m_{f,2}/m_{f,3}$. In the following, the hat (\(\,\hat{}\,\)) denotes
the division by the largest mass \(m_{f,3}\).

\paragraph{The four mass ratios parametrization}
The fact, that only two mass ratios for each fermion species are
independent parameters, gives four independent mass ratios in each
sector (quarks and leptons). An important remark at this point is, that
also four parameters are needed to fully parametrize the mixing. This
observation shall be used to build up the mixing matrix. In the standard
parametrization, those four values are three angles and one
phase---additional phases are to be rotated away by redefinition of the
fermion fields. The case of Majorana neutrinos does not allow to rotate
away the phases for the neutrinos, so two ``Majorana phases'' are
left. In the following, we will leave aside the issue of Majorana phases
and only discuss the Dirac phases. We shall show that it is possible to
use the four mass ratios of each fermion sector to entirely parametrize the
mixing without introduction of new parameters.

It is interesting to note, that a complete parametrization of the fermion
mixing in terms of the fermion mass ratios only works in the two- and
three-family case. To completely parametrize the mixing matrix, for
$n>1$ families, we need $(n-1)^2$ mixing parameters. On the other hand,
$n-1$ mass ratios are independent for each fermion species. Therefore,
only when the number of mass ratios in the corresponding fermion sector
is equal to or larger than the number of mixing parameters, $2(n-1) \geq
(n-1)^2$, this parametrization will be possible. In general, this only
works out for two or three families.

\subsection{The lower-rank approximations}
Let us investigate the effect of neglecting the first generation
masses. From now on we will work with the singular values normalized by
the largest one. In the \(\hat{m}_{f,1} \to 0\) limit, the application
of Schmidt-Mirsky's approximation theorem to the mass matrices is
consistent with the rank-two approximation. As we are neglecting all
contributions \(\mathcal{O}(\hat m_{f,1})\) we shall take into account
all corrections of the same order later on to get a more precise result
and reduce the error stemming from this approximation.

The rank two mass matrices are then given by
\begin{eqnarray}
  \hat{\cal M}_f^{r=2} =
  \left[l_{f,2} \hat{m}_{f,2} r_{f,2}^\dagger +
    l_{f,3} r_{f,3}^\dagger\right] = \begin{pmatrix}
    0 & 0 & 0 \\
    0 & \hat m^f_{22} & \hat m^f_{23} \\
    0 & \hat m^f_{32} & \hat m^f_{33}
  \end{pmatrix}.
\end{eqnarray}
In general, all the matrix elements should be different from zero.
However, it is crucial to establish a connection between a lower-rank
approximation and its origin to the Yukawa interactions. That is,
$\hat{m}_{f,1} = 0$ is equivalent to decoupling the first fermion family
from the Higgs field, $Y_{1j}^f = 0 = Y_{j1}^f$.  Effectively, thus, we
are left with a $2\times 2$ mass matrix. In the 1-1 sector, in contrast,
a phase freedom corresponding to \(\U(1)\) rotations for the
left- and right-handed fields is left, where
the second and third generation share one common phase.

Up to now, we have only used the hierarchy \(m_{f,2} \gg m_{f,1}\) to
decouple the first generation masses. According to the lower-rank
approximation theorem, the rank-two approximation differs in every
element from the full rank matrix, whereas its norm, for any chosen one,
only changes slightly. The DMHP furthermore shows \(m_{f,3} \gg
m_{f,2}\) which can be exploited to further approximate the initial mass
matrix by a rank-one matrix,
\begin{eqnarray}\label{eq:rank-1}
  \hat{\cal M}_f^{r=1} =
  l_{f,3} r_{f,3}^\dagger  =
  \begin{pmatrix}
    0 & 0 & 0 \\
    0 & 0 & 0 \\
    0 & 0 & 1
  \end{pmatrix}.
\end{eqnarray}
Successively reducing the rank of the mass matrices helps to simplify
the parametrization without loosing track of the parameters. It is,
however, not necessary to work in the very crude rank one approximation,
but sufficient to consider as a starting point the rank two approximation.

Eq.~\eqref{eq:rank-1} reveals a left-over \(\U(2)\) rotation in the
1-2 plane and one common \(\U(1)\) factor for the third generation. We
want to emphasize that the described picture of lower-rank
approximations follows what is discussed in the literature as
minimally broken flavor symmetry \cite{Barbieri:1995uv,
  Barbieri:1996ww, Barbieri:1997tu}. In the limit of vanishing Yukawa
couplings, the SM exhibits a \([\U(3)]^5\) global flavor symmetry
(\([\U(3)]^6\) if right-handed neutrinos are considered). Each
individual \(\U(3)\) flavor symmetry gets gradually broken
\[
\U(3) \;\stackrel{M_3}{\longrightarrow}\;
\U(2)\;\stackrel{M_2}{\longrightarrow}\;
\U(1)\;\stackrel{M_1}{\longrightarrow}\; \text{nothing},
\]
with \(M_3 > M_2 > M_1\) which simultaneously occurs in the up- and down
sector and trivial \(\U(1)\)s are left out for readability. After
the first symmetry breaking step at \(M_3\), one global phase freedom is
left for the third generation that is combined to a global \(\U(1)\) for
the second and third after the following symmetry breaking. There is one
residual \(\U(1)\) symmetry left for all fermions in each sector at the
end which is either baryon or lepton number. It is not only safe to work
with \(M_3 \gg M_2\)---where we are at the \(\U(2)\) flavor symmetries
of \cite{Barbieri:1995uv, Barbieri:1996ww, Barbieri:1997tu}, but even
\(M_2 \gg M_1\) which allows to work with the rank-two approximation at
a sufficiently low scale and perform the final symmetry breaking step at
say the electroweak scale.

\(\U(2)\) symmetric Yukawa couplings give a well-motivated and
frequently used setup to study flavor physics in supersymmetric
\cite{Crivellin:2008mq, Crivellin:2011sj} and unified
\cite{Barbieri:1996ww} theories and are still a viable tool to discuss
recent results in flavor physics \cite{Barbieri:2012uh,
  Buras:2012sd}. Application to lepton flavor physics was also
considered \cite{Carone:1997qg, Tanimoto:1997zw, Hall:1998cu,
  Barbieri:1999pe}, recently also in the context of \([\U(3)]^5\)
breaking \cite{Blankenburg:2012nx}. The implication of \(\U(2)\) flavor
symmetries which can be used in a weaker symmetry assignment
\cite{Aranda:1999kc}, is the arrangement of the first two families into
one doublet whereas the third family transforms as a singlet under the
flavor symmetry. This assignment can be achieved with the minimal
discrete symmetry \(S_3\) \cite{Hall:1995es, Kubo:2003iw, Morisi:2006pf,
  Feruglio:2007hi, Teshima:2011wg} that was
applied to neutrinos \cite{Jora:2009gz} as well as quarks
\cite{Canales:2013cga}.

The important point in the discussion of fermion mixings in terms of
fermion masses via lower-rank approximations is, that we implicitly
assume the maximal \([\U(3)]^6\) flavor symmetry broken with each symmetry
breaking step \emph{occurring simultaneously} for each subgroup
\([\U(3)]^6 = \U(3)_Q \times \U(3)_u \times \U(3)_d \times \U(3)_L
\times \U(3)_e \times \U(3)_\nu\).

\paragraph{Order of independent rotations}
To parametrize the three-fold mixing, we follow the commonly used three
successive rotations depending on one angle and one phase each. The
order of these transformations needs to follow the consecutive breakdown
of the initial \(\U(3)\) symmetry as implied by the hierarchy in the
masses. Therefore,
\begin{eqnarray} \label{eq:Euler-order}
  L^f = L^f_{12}(\theta_{12}^f,\delta^f_{12})
  L^f_{13}(\theta_{13}^f,\delta^f_{13})
  L^f_{23}(\theta_{23}^f,\delta^f_{23}),
\end{eqnarray}
where each individual rotation is parametrized by one angle
\(\theta^f_{ij}\) and one phase \(\delta^f_{ij}\).\footnote{Later, when
  reparametrizing the individual rotations in terms of the masses we
  will see that some of these six mixing parameters are unphysical while
  the rest can be expressed solely by two mass ratios.}

Note that this set of rotations diagonalize the mass matrices for each
fermion type. The resulting mixing matrices are the product of all the
individual rotations
\[
V_{\text{CKM}} = L^u {L^d}^\dag = L^u_{12} L^u_{13} L^u_{23}
{L^d_{23}}^\dag {L^d_{13}}^\dag {L^d_{12}}^\dag
\]
and
\[
U_{\text{PMNS}} = L^e {L^\nu}^\dag = L^e_{12} L^e_{13} L^e_{23}
{L^\nu_{23}}^\dag {L^\nu_{13}}^\dag {L^\nu_{12}}^\dag.
\]
By convention, up- and down-type rotations are exchanged for leptons.

\subsection{The effective $2 \times 2$ mass matrix}
It is instructive to first study the two-family limit in the rank-two
approximation following from \(\hat m_{f,1} \ll 1\). The second
hierarchy \(m_{f,2} \ll m_{f,3}\) implies a $2\times 2$ mass matrix of
the form
\begin{equation}\label{eq:2by2mass}
  \hat{\bf m}^f = \begin{pmatrix}
    \hat m^f_{ss} & \hat m^f_{sl} \\ \hat m^f_{ls} & \hat m^f_{ll}
  \end{pmatrix},
\end{equation}
with hierarchical elements $|\hat m^f_{ll}|^2 \gg |\hat m^f_{sl}|^2,
|\hat m^f_{ls}|^2 \gg |\hat m^f_{ss}|^2$ and where we are now
generically treating two fermion families whose singular values obey the
hierarchy, $\sigma_l \gg \sigma_s$. In general, the matrix elements are
complex numbers. The labelling \(s\) and \(l\) refers to the
corresponding smaller and larger singular value, respectively. It can be
shown that the order of magnitude of \(\hat m^f_{ss}\) is about
\(\mathcal{O}(|\hat{m}^f_{sl}|^2)\) (see~\ref{app:m11}). In the
following, we work with the approximation \(\hat m^f_{ss} = 0\).

Unlike most considerations, we take the outcome of the DMHP and
  minimal flavor symmetry breaking to set the magnitudes of the
off-diagonals equal---the phases are not constrained, such that
\[
|\hat{m}^f_{sl}| = |\hat{m}^f_{ls}|
\,\qquad\text{not}\qquad
\hat{m}^f_{sl} = (\hat{m}^f_{ls})^*,
\] as implied by the requirement of an Hermitian mass matrix.  We only
need normal mass matrices.\footnote{A matrix is normal if the left and
  right Hermitian products are the same: \(\mathbf{m}\,\mathbf{m}^\dag =
  \mathbf{m}^\dag\,\mathbf{m}\).} In both cases (normal and Hermitian),
the left and right Hermitian products are diagonalized by the same
unitary transformation. For a normal mass matrix, however, the phases
can be arranged in a way that the off-diagonal magnitudes do not have to
be the same. We only constrain the matrix of absolute values to be
symmetric, whereas the phases can be arbitrary:
\begin{equation}\label{eq:approxmassmat}
  \hat{\bf m}^f = \begin{pmatrix}
    0 & |\hat m^f_{sl}| e^{i\delta_{sl}^f} \\
    |\hat m^f_{sl}| e^{i\delta_{ls}^f} & \hat m^f_{ll}
  \end{pmatrix}.
\end{equation}

As a self-consistency check, it is important to verify that the required
hierarchy in all the mass matrix elements of the full-rank scenario
actually is respected when expressing the matrix elements in terms of
the masses (singular values).

\paragraph{Reparametrization in terms of the singular values}
Due to our lack of knowledge of right-handed flavor mixing, the relevant
object that determines our phenomenology is the Hermitian product
\({\mathbf{n}}^f = {\mathbf{m}}^f\,({\mathbf{m}}^f)^\dag\), which
exhibits two invariants: \(\tr{\mathbf{n}}^f = \sigma^{f2}_s +
\sigma^{f2}_l\) and \(\det{\mathbf{n}}^f = \sigma^{f2}_s
\sigma^{f2}_l\). The small and large singular value are denoted by
\(\sigma^f_s\) and \(\sigma^f_l\), respectively. Through means of the
two invariants, we find
\begin{eqnarray} \label{eq:twoInvariants}
|{\hat m}^f_{sl}| =
  \sqrt{\hat{\sigma}_{sl}^f}, \quad\quad {\text{and}} \quad\quad |\hat
  m^f_{ll}| = 1 - \hat{\sigma}_{sl}^f,
\end{eqnarray}
where we have expressed for a generic treatment the normalized ratio of
the small singular value over the large one as $\hat{\sigma}^f_{sl}
\equiv \sigma^f_s / \sigma^f_l$.

This reparametrization nicely shows the result of the Schmidt-Mirsky
approximation theorem: on the one hand, $|\hat m^f_{ll}|^2 \gg |\hat
m^f_{sl}|^2$, while on the other hand, $|\hat m^f_{ll}| = 1$ is the only
non-vanishing matrix element in the limit $\hat{\sigma}^f_s \rightarrow
0$.

The left unitary transformation corresponding to the diagonalization of
this matrix is given by
\begin{eqnarray}\label{eq:2diag}
  L_{sl}^f(\hat{\sigma}^f_{sl},\delta^f_{sl}) = \frac{1}{\sqrt{1+\hat{\sigma}^f_{sl}}}
  \begin{pmatrix}
    1 & e^{-i\delta^f_{sl}} \sqrt{\hat{\sigma}^f_{sl}} \\
    - e^{i\delta^f_{sl}} \sqrt{\hat{\sigma}^f_{sl}} & 1
  \end{pmatrix}.
\end{eqnarray}
This result has been already discussed previously by many
authors~\cite{Fritzsch:1977za, Weinberg:1977hb, Raby:1995uv,
  Fritzsch:1999ee}. The mixing angle can be obtained from
\(\tan\theta_{sl}^f = \sqrt{\hat\sigma^f_{sl}}\).\footnote{Another
  solution can be found, that behaves wrongly in the limit
  \(\hat{\sigma}^f_{sl} \to 0\) and gives maximal mixing
  \(\tan\theta^f_{sl} \to \infty\) instead of zero mixing.} Note that
this relation indeed is the Gatto-Sartori-Tonin result,
see~Eq.\eqref{eq:GSTrel}.

\paragraph{The two-family mixing matrix}
Eq.~\eqref{eq:2diag} diagonalizes the mass matrix of one fermion
type. In the weak charged current, an $a$-type fermion ($a = u,e$) meets
a $b$-type fermion ($b=d,\nu$), so we need two such diagonalizations to
describe fermion mixing in the charged current interactions. Anyway, two
unitary \(2\times 2\) rotations do not commute, and the new mixing
parameters are not just the sum or difference of the former ones:
\(\theta_{sl} \neq \theta^a_{sl} \pm \theta^b_{sl}\) and \(\delta \neq
\delta^a_{sl} \pm \delta^b_{sl}\). Explicitly,
\begin{equation}\label{eq:twofold}
  V_{sl} = L_{sl}^a {L_{sl}^b}^\dag = \diag(1, e^{-i\delta_{sl}^a})
  \begin{pmatrix}
    \sqrt{1 - \lambda^2} e^{-i\delta_0} & \lambda e^{-i\delta} \\
    - \lambda e^{i\delta} & \sqrt{1 - \lambda^2} e^{i\delta_0}
  \end{pmatrix} \diag(1, e^{i \delta_{sl}^a}),
\end{equation}
where we factored out the phase \(\delta_{sl}^a\). This choice is
completely arbitrary, the same is true for \(\delta_{sl}^b\). The
relevant phases \emph{inside} the matrix only depend on the
\emph{difference}. The mixing can then be obtained in the following way
\begin{align}\label{eq:theta}
  \lambda =\sin\theta_{sl} &= \sqrt{\frac{\hat{\sigma}^a_{sl} +
      \hat{\sigma}^b_{sl} - 2 \sqrt{ \hat{\sigma}^a_{sl}
        \hat{\sigma}^b_{sl}} \cos(\delta^a_{sl} -
      \delta^b_{sl})}{(1+\hat{\sigma}^a_{sl})(1+\hat{\sigma}^b_{sl})}}
  , \\
  \tan\delta &= \frac{\hat{\sigma}^b_{sl} \sin(\delta^a_{sl} -
    \delta^b_{sl})}{\hat{\sigma}^a_{sl} - \hat{\sigma}^b_{sl}
    \cos(\delta^a_{sl} -
    \delta^b_{sl})}, \\
  \tan\delta_0 &= \frac{\hat{\sigma}^a_{sl} \hat{\sigma}^b_{sl}
    \sin(\delta^a_{sl} - \delta^b_{sl})}{1 +\hat{\sigma}^a_{sl}
    \hat{\sigma}^b_{sl} \cos(\delta^a_{sl} - \delta^b_{sl})}.
\end{align}
The functional dependence on the two initial complex phases is found to
be only their difference. From the hierarchies \(\hat\sigma^x_{sl} =
\sigma^x_s / \sigma^x_l \ll 1\) (for \(x=a,b\)) follow the new phases to
be approximately given by \(\tan\delta \approx - \tan(\delta_{sl}^a -
\delta_{sl}^b)\) and \(\tan\delta_0 \approx 0\). For the full-rank
scenario, however, this simple conclusion cannot be drawn---it actually
holds for the ``initial'' 2-3 rotation, but not anymore when subsequent
rotations are added.

\paragraph{Comment on the complex phases}
In general, the complex phases of the initial mass matrix elements are
not constrained to a particular value. The employed matrix invariants
only restrict the moduli of the matrix elements, the phases are
unconstrained. There is nevertheless an ambiguity in those phases that
is not necessary to set up a full parametrization of fermion mixing in
the SM. The standard parametrization uses three successive rotations
with $\theta_{ij} \in [0,\frac{\pi}{2}]$ and one complex phase
$\delta_{CP} \in [0,2\pi)$. These four parameters are sufficient to
describe both mixing and CP violation in each fermion sector (unless we
want to include a description of Majorana phases for neutrinos). In
contrast, we have four mass ratios---and the freedom to put either real
or purely imaginary matrix elements. This last choice can be achieved by
restricting all phases to be either maximal CP violating (\(\pi/2\) or
\(3\pi/2\)) or CP conserving (\(0\) or \(\pi\)). Interestingly, at the
end, there is no freedom in phase choices at all and we find that
only the 1-2 phase is allowed to be maximally CP violating, which indeed
follows from a symmetry argument.

\subsection{The full-rank picture}
Working in the lower-rank approximations, we are neglecting the first
generation mass (\(\hat{m}_{f,1} = 0\)) in the 2-3 rotation and the
second generation mass (\(\hat{m}_{f,2} = 0\)) while performing the 1-3
rotation. The last transformation that appears in
Eq.~\eqref{eq:Euler-order} acting in the 1-2 plane needs no
approximation. It affects only the upper left \(2\times 2\) submatrix
and is an exact diagonalization. In all cases, the mass matrices are of
the form \eqref{eq:2by2mass}, where the elements are properly
distributed over the \(3\times 3\) matrix elements. All residual matrix
elements are zero. The same holds for the arising rotation matrices that
are \(3\times 3\) generalizations of Eq.~\eqref{eq:2diag}.

Working in the leading order approximations shows a subtle
inconsistency: neglecting \(\mathcal{O}(\hat m_{f,2})\) terms in the 1-3
rotation means actually ignoring a large effect, because
\(\mathcal{O}(\hat m_{f,1}) = \mathcal{O}(\hat m_{f,2}^2)\). Moreover, to
include \(\mathcal{O}(\hat m_{f,1})\) contributions in the 1-3
rotation following the initial rotation in the 2-3 plane, we first have to
consider contributions of the same order that were missing in the
initial rotation. Therefore, we briefly discuss how to consistently
include corrections of missing pieces to improve the result.

\paragraph{Inclusion of corrections}
We include the corrections as correcting (small) rotations. This
procedure is crucial in view of the symmetry breaking chain from an
enhanced flavor symmetry, as \([\U(3)]^3\) (corresponding to a rank-zero
mass matrix), down to the least symmetry left over. Since each breaking
step is done by a small parameter, we do not disturb much by adding
perturbations. Moreover, by repeatedly applying rotations, this
guarantees from the very beginning normalized eigenvectors, and
furthermore, an inclusion of formally higher order terms in perturbation
theory. This can be seen from the following example of two real
rotations, where \(\hat\epsilon \lesssim
\hat{\sigma}^f_{sl}\):\footnote{The two signs reflect the freedom of
  choice for a clockwise or counterclockwise correcting rotation.}
\begin{eqnarray} {L^{f}_{sl}}^{(p=1)} = {L^{f}_{sl}}^{(1)}(\pm\hat
  \epsilon){L^{f}_{sl}}^{(0)} (\hat{\sigma}^f_{sl}) =
  \begin{pmatrix} \cos{\theta^f_{sl}}^{(p=1)} &
    \sin{\theta^f_{sl}}^{(p=1)} \\ -\sin{\theta^f_{sl}}^{(p=1)}&
    \cos{\theta^f_{sl}}^{(p=1)}
  \end{pmatrix},
\end{eqnarray}
and the new angle is given by
\begin{eqnarray} \sin{\theta^f_{sl}}^{(p=1)} =
  \frac{\sqrt{\hat{\sigma}_{sl}^{f}} \pm \sqrt{\hat
      \epsilon}}{\sqrt{(1+\hat{\sigma}_{sl}^{f})(1+\hat\epsilon)}}.
\end{eqnarray}
For real rotations, the requirement \(\hat\epsilon \lesssim
\hat{\sigma}^f_{sl}\) is irrelevant, because \(\O(2)\) rotations
commute. Therefore, there is also no need to specify any order in the
addition of correcting rotations in each \(i\)-\(j\) plane.

Inverting this procedure shows that it is equivalent to add the
perturbation term
\begin{eqnarray}\label{eq:perturb}
  -\sqrt{\hat\epsilon} \left[ 1+(\hat{\sigma}^f_{sl})^2 -2\hat{\sigma}^f_{sl}
    + \sqrt{\hat {\epsilon}\hat{\sigma}^f_{sl} }
    (\hat{\sigma}^f_{sl}-1)\right](1-\hat \epsilon)
\end{eqnarray}
to the off diagonal matrix elements $s$-$l$ and $l$-$s$.

Continuing this, an arbitrary number of correcting rotations could be
added in each \(2\times 2\) rotation:
\begin{eqnarray}\label{eq:theta-pert}
  \sin{\theta^f_{sl}}^{(p=n)} =
  \frac{\sum_{j=0}^{n}(-1)^{\delta_j}\sqrt{\hat{a}_j} +
    \mathcal{O}\left(\left[\sqrt{\hat a_i \hat a_j \hat a_k}\right]_{i
        \neq j \neq k}\right)}
  {\sqrt{(1+\hat{a}_{0})(1+\hat{a}_1)(1+\hat{a}_2) \cdots  (1+\hat{a}_n)}},
\end{eqnarray}
where we have denoted $\hat{a}_0 \equiv \hat{\sigma}_{sl}^f$ and \(\hat
a_{i>0}\) for the parameters of the following rotations. Each
$(-1)^{\delta_i}$ is the orientation of the \(i\)-th rotation, which is
either clockwise or counterclockwise (plus or minus). We neglect in
Eq.~\eqref{eq:theta-pert} all trilinear and higher products of \(\hat
a_i\), where no \(\hat a_i^2\) and no even products appear. Let us
emphasize here, nevertheless, that these correcting rotations do not
follow the traditional procedure of perturbation theory where we could
naively think that the following new correcting rotation is a power of
the previous one. Inclusion of new correcting rotations requires a
careful treatment.  We have found to be sufficient to include two
correcting rotations to the mixing matrix parametrization which are the
contributions \(\mathcal{O}(\hat m_{f,1})\), \(\mathcal{O}(\hat
m_{f,2}^2)\), and \(\mathcal{O}( \hat m_{f,1} \cdot \hat m_{f,2} )\)
which are of the same order as the neglected terms in each case.

\paragraph{First rotation: The 2-3 sector}
Starting from the rank-two approximation, we loose track of all
\(\sqrt{\hat{m}_{f,1}}\) contributions in the mass matrix. However, all
correcting rotations have to be consistent with the initial
approximation (\(\hat{m}_{f,1} \to 0\)) and, moreover, all ``higher
order'' contributions (\(\sim\hat{m}_{f,2}^2\), \(\sim\hat{m}_{f,1}^2\))
are already covered as can be seen from \eqref{eq:perturb}. We therefore
conclude, that all reasonable rotations in the 2-3 plane can be
expressed as
\begin{eqnarray}
  {L^f_{23}}^{(p=2)} ={L^{f}_{23}}^{(2)}(\hat{m}_{f,1}\cdot\hat{m}_{f,2})
  {L^{f}_{23}}^{(1)}(\hat{m}_{f,1}){L^{f}_{23}}^{(0)} (\hat{m}_{f,2}).
\end{eqnarray}
Additionally, in principle, there is a freedom in the choice of the
complex phase, which can be boiled down to the two different sign
choices.

\paragraph{Second rotation: The 1-3 sector}
What follows is the same procedure in the 1-3 sector after the 2-3
rotations have been done. In this case, the $p=2$ leading correcting
rotations are
\begin{eqnarray}
  {L^f_{13}}^{(p=2)} =
  {L^{f}_{13}}^{(2)}(\hat{m}_{f,1}\cdot\hat{m}_{f,2}) {L^{f}_{13}}^{(1)}(\hat{m}_{f,2}^2)
  {L^{f}_{13}}^{(0)} (\hat{m}_{f,1}).
\end{eqnarray}

\paragraph{Last rotation: The 1-2 sector}
No approximation is left anymore, therefore the exact rotation is
expressed as
\begin{eqnarray}
  {L^f_{12}} = L^f_{12} (\frac{\hat{m}_{f,1}}{\hat{m}_{f,2}}, \delta^f_{12}),
\end{eqnarray}
where we now explicitly put the phase \(\delta^f_{12}\). This occurrence
is very clear from the rank evolution: in the rank-one approximation,
there is the freedom of a \(\U(2)\) rotation left in the 1-2 block. The
initial 2-3 and 1-3 rotations can always be taken real, the only
possible phase then sits in the 1-2 rotation.

The necessity of correcting rotations is very apparent from the flavor
symmetry breaking chain: First, in the rank-two approximation we have
\[
\begin{pmatrix} 0 & 0 & 0 \\ 0 & X & X \\ 0 & X & X \end{pmatrix}
\stackrel{L_{23}^{(0)}}{\longrightarrow}
\begin{pmatrix} 0 & 0 & 0 \\ 0 & X & 0 \\ 0 & 0 & X \end{pmatrix}.
\]
After performing the symmetry breaking step to the full-rank matrix, we
get contributions in all matrix elements not larger than
\(\mathcal{O}(\sqrt{\hat{m}_{f,1}})\)---also in off-diagonal components
that were already rotated away:
\[
\begin{pmatrix} * & * & * \\ * & X & * \\ * & * & X \end{pmatrix}.
\]
So, we indeed have to consider higher order corrections to the initial
rotation. The correcting rotations also do not spoil the required
hierarchy. After the successive 2-3 and 1-3 rotations there is a
contribution shuffled into the 1-1 entry which is \(\sim s_{13}^2 m_{33}
\sim \mathcal{O}(\hat{m}_{f,1})\) and therefore of higher order compared
to \(\mathcal{O}(\hat{m}_1/\hat{m}_2)\), the original 1-1 element.

\section{Applying the DMHP to phenomenology}
\label{sec:applpheno}
By building up the mixing matrices following the procedure of the
previous section, there appears the impression of an arbitrariness in
the choice of complex phases. This arbitrariness can be attenuated taking
into account some well motivated considerations. First, complex phases
appear pairwise in the up- and down-type fermion sectors. We therefore
have the freedom to keep track of them in only one sector and set all
phases in the other one equal to zero. The charged current mixing matrix
is therefore constructed in the following way:
\begin{align}
V_\text{CKM} &= L^u {L^d}^\dag, \nonumber\\
L^u &= L^u_{12}\left(\frac{m_u}{m_c}\right)
L^u_{13}\left(\frac{m_um_c}{m_t^2}\right)
L^u_{13}\left(\frac{m_c^2}{m_t^2}\right)
L^u_{13}\left(\frac{m_u}{m_t}\right)
L^u_{23}\left(\frac{m_um_c}{m_t^2}\right)
L^u_{23}\left(\frac{m_u}{m_t}\right)
L^u_{23}\left(\frac{m_c}{m_t}\right), \\
{L^d}^\dag &= {L^d_{23}}^\dag\left(\frac{m_s}{m_b}, \delta^{(0)}_{23}\right)
{L^d_{23}}^\dag\left(\frac{m_d}{m_b}, \delta^{(1)}_{23}\right)
{L^d_{23}}^\dag\left(\frac{m_dm_s}{m_b^2}, \delta^{(2)}_{23}\right)
\;\times \nonumber \\ &\quad
{L^d_{13}}^\dag\left(\frac{m_d}{m_b}, \delta^{(0)}_{13}\right)
{L^d_{13}}^\dag\left(\frac{m_s^2}{m_b^2}, \delta^{(1)}_{13}\right)
{L^d_{13}}^\dag\left(\frac{m_dm_s}{m_b^2}, \delta^{(2)}_{13}\right)
{L^d_{12}}^\dag\left(\frac{m_d}{m_s}, \delta_{12}\right),
\label{eq:down} \\
U_\text{PMNS} &= L^e {L^\nu}^\dag, \nonumber \\
L^e &= L^e_{12}\left(\frac{m_e}{m_\mu}\right)
L^e_{13}\left(\frac{m_\mu^2}{m_\tau^2}\right)
L^e_{13}\left(\frac{m_em_\mu}{m_\tau^2}\right)
L^e_{13}\left(\frac{m_e}{m_\tau}\right)
L^e_{23}\left(\frac{m_em_\mu}{m_\tau^2}\right)
L^e_{23}\left(\frac{m_e}{m_\tau}\right)
L^e_{23}\left(\frac{m_\mu}{m_\tau}\right), \\
{L^\nu}^\dag &= {L^\nu_{23}}^\dag\left(\frac{m_{\nu2}}{m_{\nu3}}, \delta^{(0)}_{23}\right)
{L^\nu_{23}}^\dag\left(\frac{m_{\nu1}}{m_{\nu3}}, \delta^{(1)}_{23}\right)
{L^\nu_{23}}^\dag\left(\frac{m_{\nu1}m_{\nu2}}{m_{\nu3}^2}, \delta^{(2)}_{23}\right)
\;\times \nonumber \\ &\quad
{L^\nu_{13}}^\dag\left(\frac{m_{\nu1}}{m_{\nu3}}, \delta^{(0)}_{13}\right)
{L^\nu_{13}}^\dag\left(\frac{m_{\nu2}^2}{m_{\nu3}^2}, \delta^{(1)}_{13}\right)
{L^\nu_{13}}^\dag\left(\frac{m_{\nu1}m_{\nu2}}{m_{\nu3}^2}, \delta^{(2)}_{13}\right)
{L^\nu_{12}}^\dag\left(\frac{m_{\nu1}}{m_{\nu2}}, \delta_{12}\right).
\label{eq:neutrino}
\end{align}

\begin{figure}[tb]
\begin{minipage}{.5\textwidth}
\includegraphics[width=\textwidth]{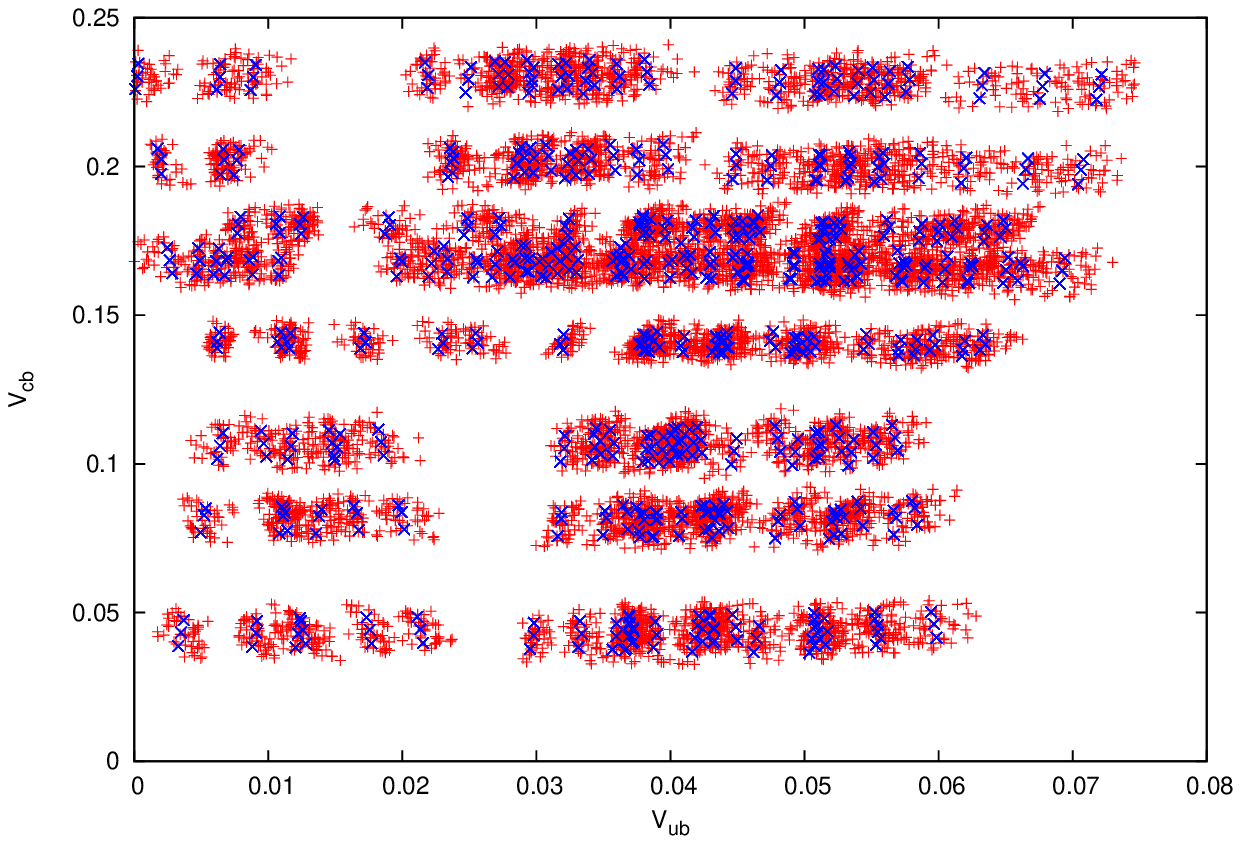}
\end{minipage}%
\begin{minipage}{.5\textwidth}
\includegraphics[width=\textwidth]{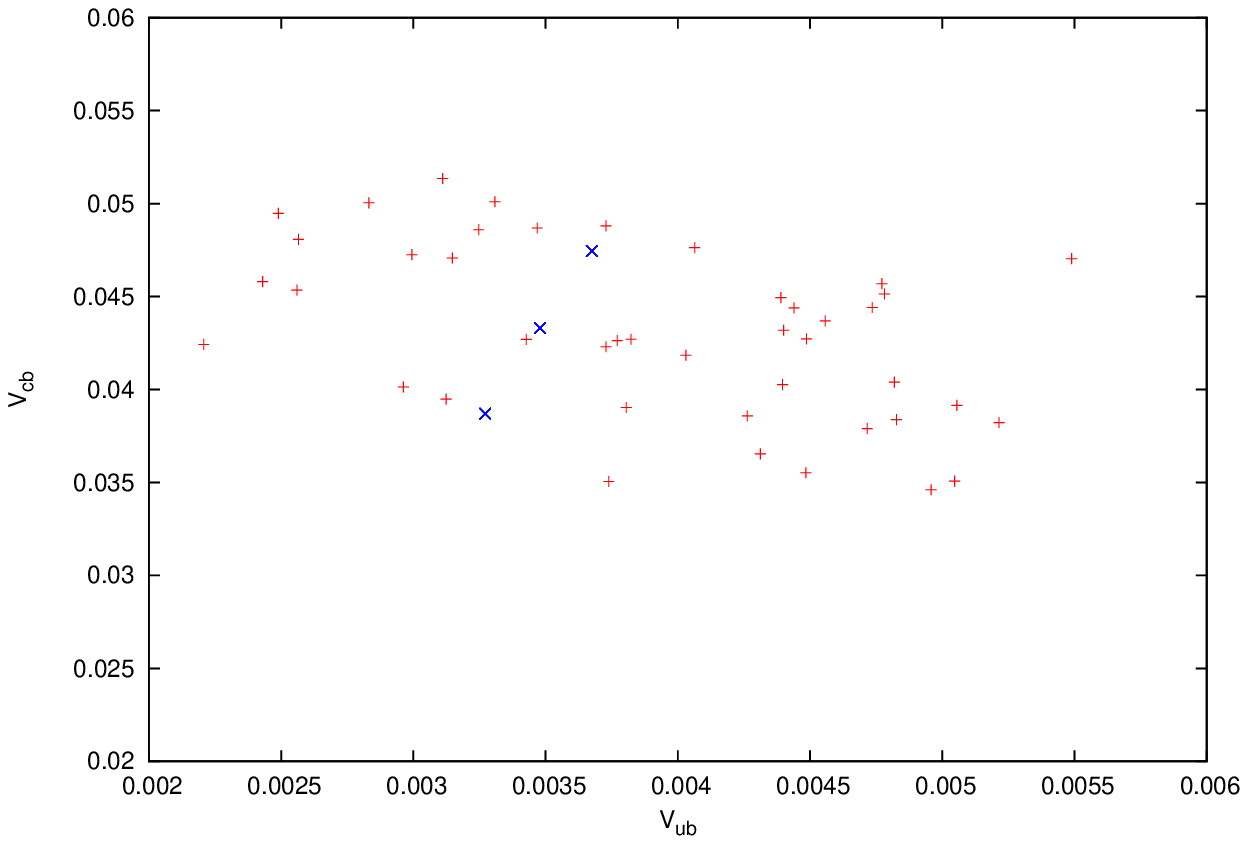}
\end{minipage}
\caption{Distribution of allowed values in the \(V_{ub}\)-\(V_{cb}\)
  plane. The small red points show allowed regions where the masses were
  varied in their \(1\,\sigma\) regimes, the blue crosses show the
  values coming from the central values of the masses. Right: zoom into
  the phenomenological viable region. There are only three distinct
  phase choices leading to both small values for \(V_{ub}\) and
  \(V_{cb}\).}
\label{fig:CKMphases}
\end{figure}

\begin{figure}[tb]
\begin{minipage}{.5\textwidth}
\includegraphics[width=\textwidth]{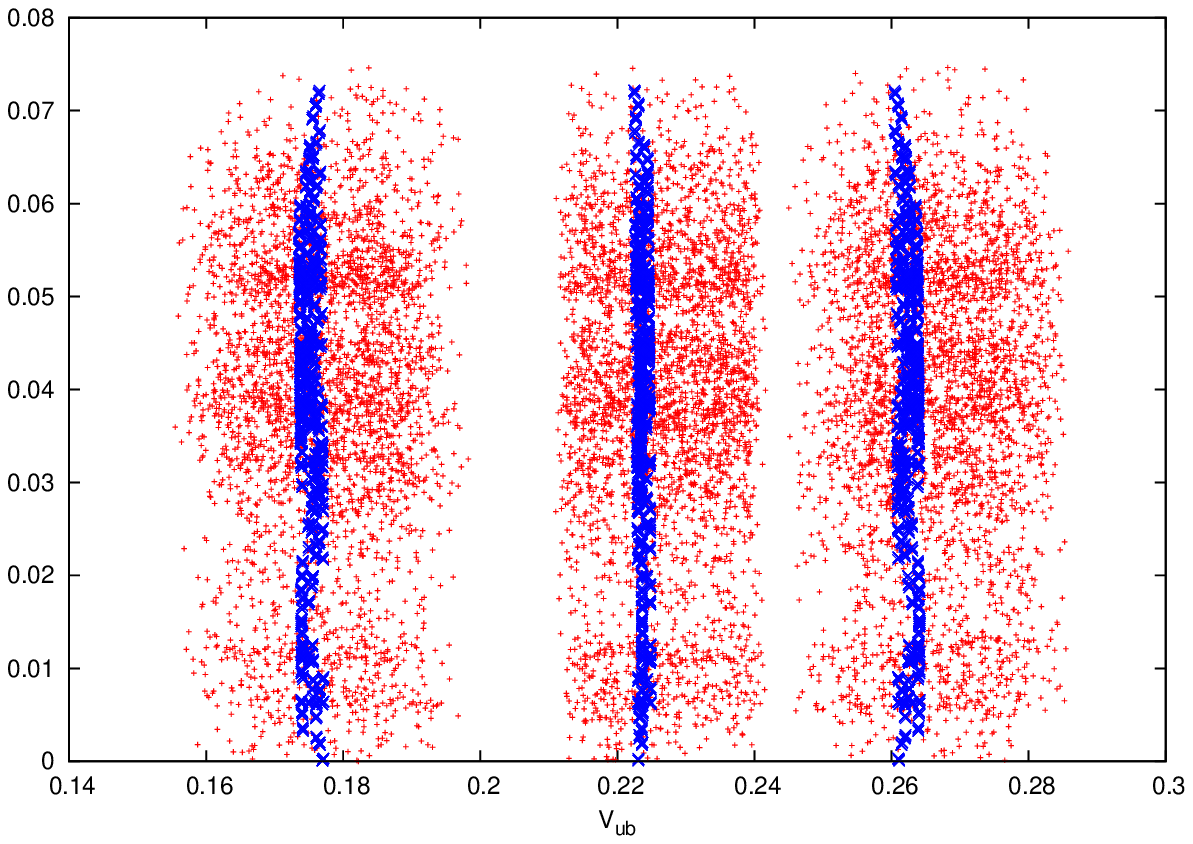}
\end{minipage}%
\begin{minipage}{.5\textwidth}
\includegraphics[width=\textwidth]{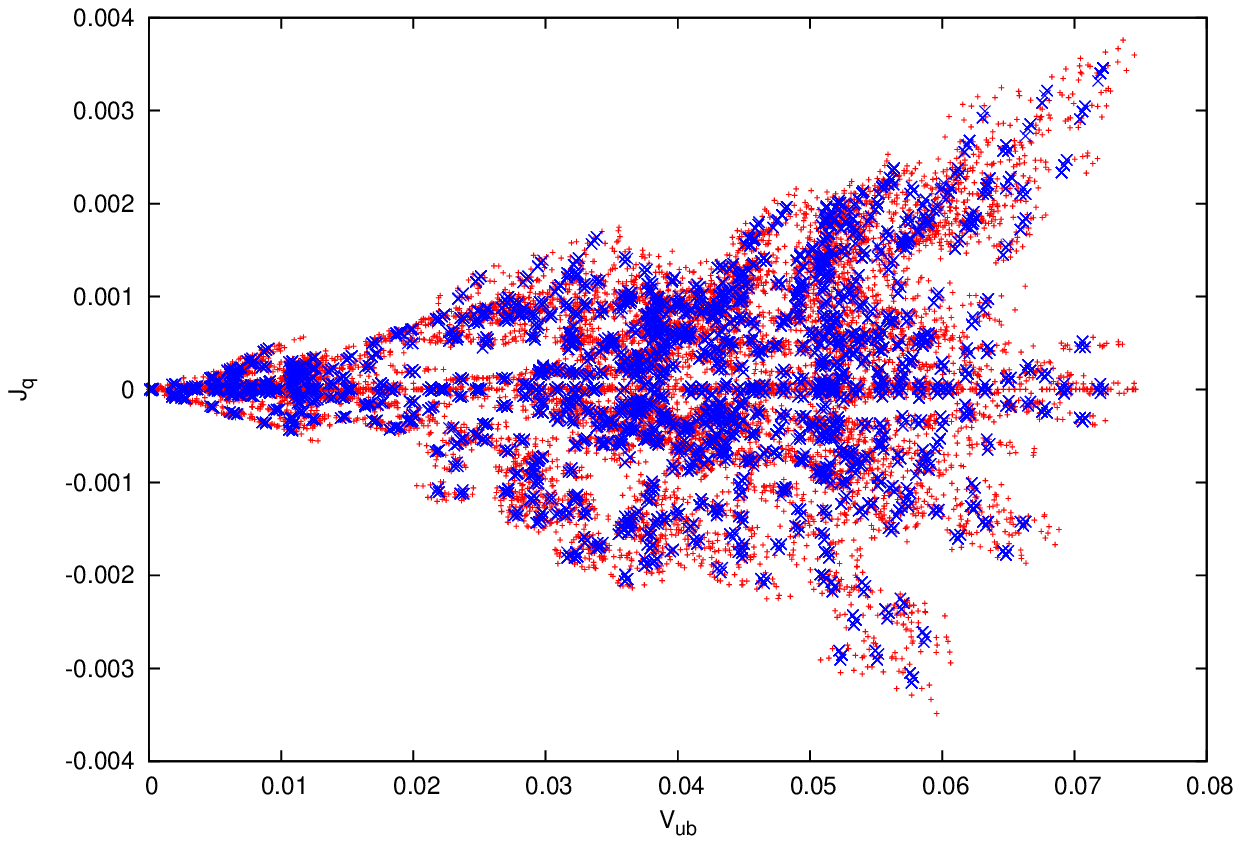}
\end{minipage}
\caption{The left plot shows the regions for the Cabibbo angle (more
  exactly \(V_{us}\)---clearly the solution \(\delta_{12} =
  \frac{\pi}{2}\) is favoured (which corresponds to the stripe in the
  center). On the right side, the rephasing invariant \(J_q\) is shown
  against \(V_{ub}\). Color code as in Fig.~\ref{fig:CKMphases}: red
  dots are points with masses varied in the \(1\,\sigma\) regimes, blue
  crosses are the central values.}
\label{fig:cabibbo_jarlskog}
\end{figure}

The method itself is not quite arbitrary at all. For the CKM mixing it
gives well-separated regions that have to be entered with a specific
choice for the phases (see Fig.~\ref{fig:CKMphases}). Since both quark
masses as well as CKM mixing matrix entries are rather well measured,
this observations allows us to set the phases. We find only one distinct
choice. Moreover, we make a \emph{minimal} choice: on the one hand, we
allow CP phases to be either maximally CP violating or CP conserving. On
the other hand, we find, that the only maximally CP violating phase has
to be in the 1-2 rotation of the down-type quarks or neutrinos,
respectively. This can be seen from Fig.~\ref{fig:cabibbo_jarlskog}
where the three bands correspond to a phase \(\delta_{12} = 0,
\frac{\pi}{2} \text{ and } \pi\).

The previously derived subsequent rotations only depend on four mass
ratios in each fermion sector and have to be faced with phenomenological
data. As input values we are using the quark and lepton masses only (see
~\ref{app:data}) and then give a prediction for the neutrino masses
to be in agreement with observations of neutrino mixing in this setup.

\subsection{Minimal or maximal CP violation}
The nature of the complex phases and its impact in the mixing matrix
elements needs further investigation. Giving a solution to this problem
is, however, outside the scope of this work. We shall use our
observation to distribute the CP violating phase properly and leave the
origin of CP violation for later work.

A final comment can be done, though, that guarantees the uniqueness of
the parametrization. In Fig.~\ref{fig:CKMphases}, we show the maximally
allowed ranges for the mixing matrix elements \(V_{ub}\) and
\(V_{cb}\). The amount of data points was constructed choosing the quark
masses from their \(1\,\sigma\) regimes and randomly taking every phase in
the final paramerization from the set \(\lbrace 0, \frac{\pi}{2},
\pi\rbrace\). It is sufficient to constrain oneself to this set which
gives the minimal and maximal allowed amount of CP
violation~\cite{Masina:2006ad}---and connected to that minimal and
maximal mixing. The latter can be seen from Eq.~\eqref{eq:theta} for the
two-generation sub-case: the phase difference \(\delta^a_{sl} -
\delta^b_{sl}\) controls the magnitude of the mixing angle between
minimal (\(\delta^a_{sl} - \delta^b_{sl} = 0\)) and maximal (\(\pi\))
mixing.

The fact, that only \emph{one} combination of phases survives, is
astonishing: note that all possible combinations in Eq.~\eqref{eq:down}
are generically \(3^7 = 2187\) choosing from \(\lbrace 0, \frac{\pi}{2},
\pi\rbrace\). Still, after taking \(\delta_{12} = \frac{\pi}{2}\) and
constraining the remnant phases to be either zero or \(\pi\), 64
combinations are left. It is therefore not \emph{a priori} clear that
the mass ratios alone give the right mixing. The functional dependence
on the mass ratios, however, is unique once the phases are set. We
therefore use this description to determine the position of the maximal
CP phase, where in contrast the other phases give relative minus
signs. The maximal CP violating phase in the neutrino 1-2 mixing is
somehwat different to what was found in connection with maximal
atmospheric mixing~\cite{Masina:2005hf}.

\begin{table}
\caption{The choice of phases in Eqs.~\eqref{eq:down} and
  \eqref{eq:neutrino} leading to the mixing matrices shown in
  \eqref{eq:postCKM} and \eqref{eq:prePMNS}.}
\label{tab:phases}
\centering
\begin{tabular}{cccccccc}
  \toprule
  & \(\delta_{12}\) & \(\delta_{13}^{(0)}\) & \(\delta_{13}^{(1)}\) & \(\delta_{13}^{(2)}\) & \(\delta_{23}^{(0)}\) & \(\delta_{23}^{(1)}\) & \(\delta_{23}^{(2)}\)\\
  \midrule
  CKM  & \(\frac{\pi}{2}\) & 0 & $\pi$ & $\pi$ & $0$ & $\pi$ & $\pi$ \\
  PMNS & \(\frac{\pi}{2}\) & 0 & $\pi$ & $\pi$ & $\pi$ & $\pi$ & $0$ \\
  \bottomrule
\end{tabular}
\end{table}

\subsection{Projected values of $V_{\text{CKM}}^{\text{th}}$ and $J_q$}
Consideration of all the aforementioned prescriptions gives the
following numbers for the magnitude of the mixing matrix elements (see
\ref{app:Formulae} for the explicit formulae of the mixing angles
and the Jarlskog invariant),
\begin{eqnarray}\label{eq:postCKM}
  |V_{\text{CKM}}^{\text{th}}| =
  \begin{pmatrix}
    0.974^{+0.004}_{-0.003} & 0.225^{+0.016}_{-0.011} & 0.0031^{+0.0018}_{-0.0015} \\
    0.225^{+0.016}_{-0.011} & 0.974^{+0.004}_{-0.003} & 0.039^{+0.005}_{-0.004} \\
    0.0087^{+0.0010}_{-0.0008} & 0.038^{+0.004}_{-0.004} & 0.9992^{+0.0002}_{-0.0001}
  \end{pmatrix}
\end{eqnarray}
and the following amount of CP violation as measured by the Jarlskog
invariant,
\begin{eqnarray}
  J_q = \Im(V_{us}V_{cb}V_{ub}^*V_{cs}^*) = ( 2.6^{+1.3}_{-1.0} ) \times 10^{-5},
\end{eqnarray}
where all quantities here are seen to be in quite good agreement within
the errors compared to the global fit result given by the PDG
2014~\cite{Agashe:2014kda} (see~\ref{app:data} for present
knowledge on masses and mixings). Note that generically, the amount of
CP violation is much larger (Fig.~\ref{fig:cabibbo_jarlskog}) and a
small value of \(V_{ub}\) is connected to a small \(J_q\), as expected.

\subsection{Lepton sector}
\label{sec:lept}
Quark masses show a very strong hierarchy. Charged lepton masses also
do. Neutrinos, though, do not do. Is it really viable to apply the DMHP
also to lepton mixing? Leptonic mixing angles are large, this
observation may hint to a different mechanism. However, mass ratios for
neutrinos are also large. The parametrization of fermion mixing in terms
of mass ratios allows to also cope with large mixings by large mass
ratios.  Nevertheless, we have to include a solid examination of the
errors in this approximation and see whether the same procedure as for
quarks is viable also for leptons.

Are neutrino masses hierarchical? Neither the quasidegenerate solution nor
the strong hierarchy are excluded yet. A hierarchical mass spectrum in
any case predicts a very light lightest neutrino (it still can be
exactly massless---in this case we would only have a rank two mass
matrix), where degenerate masses are likely to be tested in the near
future.

The power of the mixing parametrization in terms of mass ratios lies in
its invertibility: the formulae give us a unique description of the
missing mass ratio once the mixing angle is measured. The pattern of
neutrino masses brings us into the comfortable situation of nearly
disentangling the 1-2 from the 2-3 mixing, because \(\Delta m_{21}^2 /
\Delta m_{31}^2 \ll 1\). Additionally, the 1-2 mixing angle has the
smallest error in the global fit.

\paragraph{Predicted neutrino masses}
We do not focus on a specific model behind the theory of neutrino
masses. It is sufficient to consider an effective neutrino mass matrix
irrespective of the UV completion behind. To embed our description into
a theory of neutrino flavor, it definitely matters if neutrinos are
Dirac or Majorana. The size of the masses, however, allows to neglect RG
running in any case. Therefore, we also ignore the nature of the
neutrino mass operator. Since we take the magnitudes of the Dirac masses
symmetric for quarks, the only difference would be the off-diagonal
phase. Having this similarity in mind, the 1-2 approximation for
neutrinos follows directly from Eq.~\eqref{eq:twofold} and the
determining equation for the missing mass ratio from
Eq.~\eqref{eq:theta} with obvious relabelings:
\begin{equation}\label{eq:Ue2}
  |U_{e2}| \approx \sqrt{\frac{\hat{m}_{e\mu} +
      \hat{m}_{\nu 12} - 2 \sqrt{ \hat{m}_{e\mu} \hat{m}_{\nu 12}} \cos(\delta^e_{12} -
      \delta^\nu_{12})}{(1+\hat{m}_{e\mu})(1+\hat{m}_{\nu 12})}},
\end{equation}
where the mass ratios are \(\hat m_{e\mu} = m_e / m_\mu\) and \(\hat
m_{\nu12} = m_{\nu1} / m_{\nu2}\). The three individual neutrino
masses\footnote{We are implicitly assuming normal ordering. Inverted
  ordering is excluded by construction because it is not hierarchical in
  the minimal flavor symmetry breaking chain.} are obtained via the mass
squared differences:
\begin{equation} \label{eq:neutrino-mas-def}
\begin{aligned}
m_{\nu2} &= \sqrt{ \Delta m_{21}^2 / ( 1 - \hat{m}^2_{\nu12} ) }, \\
m_{\nu1} &= \sqrt{ m_{\nu2}^2 - \Delta m_{21}^2}, \\
m_{\nu3} &= \sqrt{\Delta m_{31}^2 - \Delta m_{21}^2 + m_{\nu2}^2}.
\end{aligned}
\end{equation}
In Eq.~\eqref{eq:Ue2}, there appears the phase difference
\(\delta^e_{12} - \delta^\nu_{12}\). Although a twofold rotation shows
no CP violation, this phase has to be considered because it appears last
in the order of successive rotations. Moreover, we observed a maximal CP
phase in the quark 1-2 sector. Albeit there is no connection between
quark and lepton mixing at this stage, we shall keep the assignment
\(\delta^e_{12} - \delta^\nu_{12} = \frac{\pi}{2}\) and get
\begin{equation}\label{eq:numassratio}
\hat{m}_{\nu1} = \frac{|U_{e2}|^2 ( 1 + \hat m_e ) - \hat m_e}{1 -
  |U_{e2}|^2 ( 1 + \hat m_e)} =
0.41 \ldots 0.45
\end{equation}
using \(\hat m_e = 0.00474\) and \(|U_{e2}| = \sin\theta_{12} = 0.54
\ldots 0.56\). The masses are calculated as
\begin{align*}
m_{\nu1} &= ( 0.0041 \pm 0.0015 )\,\eV, \\
m_{\nu2} &= ( 0.0096 \pm 0.0005 )\,\eV, \\
m_{\nu3} &= ( 0.050 \pm 0.001 )\,\eV.
\end{align*}
The errors were propagated from the \(\Delta m^2\) and added linearly to
be more conservative. Within \(3\,\sigma\), the lightest neutrino can be
massless. This prediction, however, will significantly improve with the
improved errors on \(\Delta m_{21}^2\).

The minimally and maximally allowed neutrino masses (corresponding to
\(\delta^e_{12} - \delta^\nu_{12} = 0, \pi\)) are very close:
\begin{center}
\begin{tabular}{ll}
\toprule
min (in \(\eV\)) & max (in \(\eV\)) \\
\midrule
\(m_{\nu1} = 0.0029 \pm 0.0017\) & \(m_{\nu1} = 0.0062 \pm 0.0017\) \\
\(m_{\nu2} = 0.0091 \pm 0.0003\) & \(m_{\nu2} = 0.011 \pm 0.001\) \\
\(m_{\nu3} = 0.050 \pm 0.001\) & \(m_{\nu3} = 0.050 \pm 0.001\) \\
\bottomrule
\end{tabular}
\end{center}
In any case, the lightest neutrino is much lighter than \(0.01\,\eV\).

\begin{figure}[tb]
\centering
\begin{minipage}{.5\textwidth}
\includegraphics[width=\textwidth]{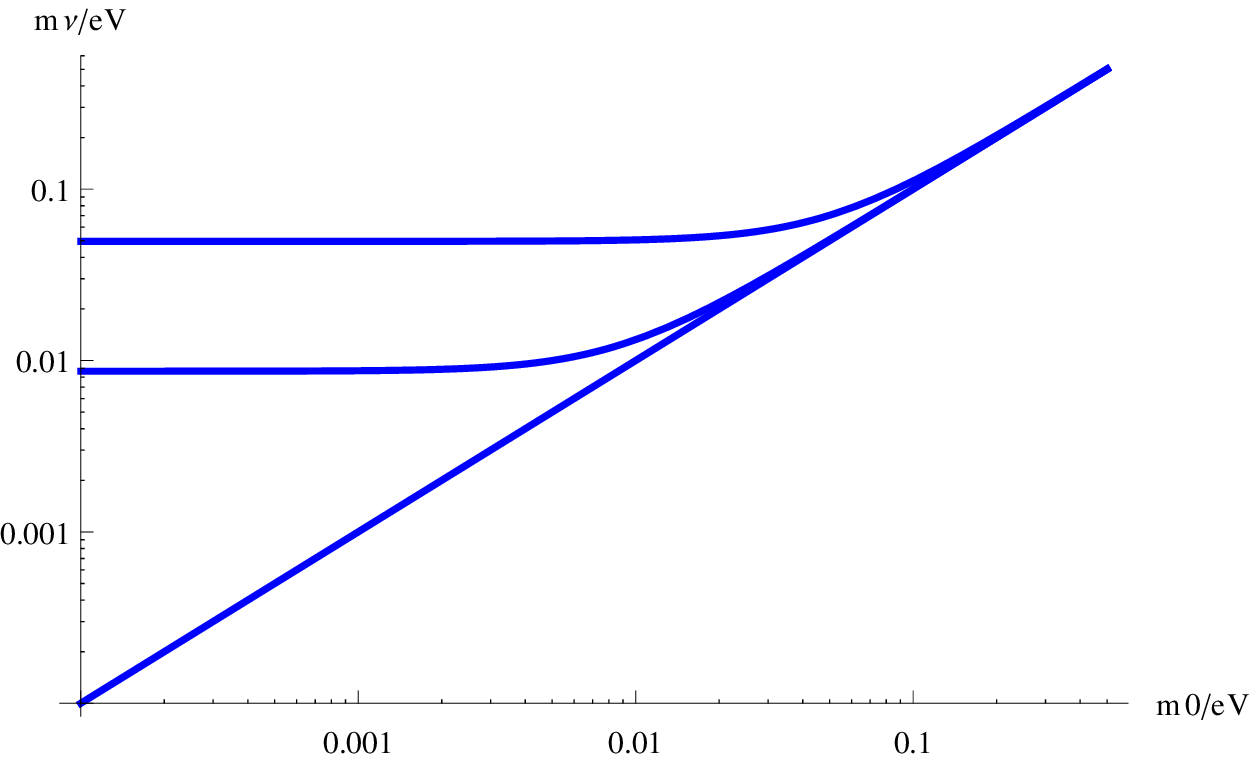}
\end{minipage}%
\begin{minipage}{.5\textwidth}
\includegraphics[width=\textwidth]{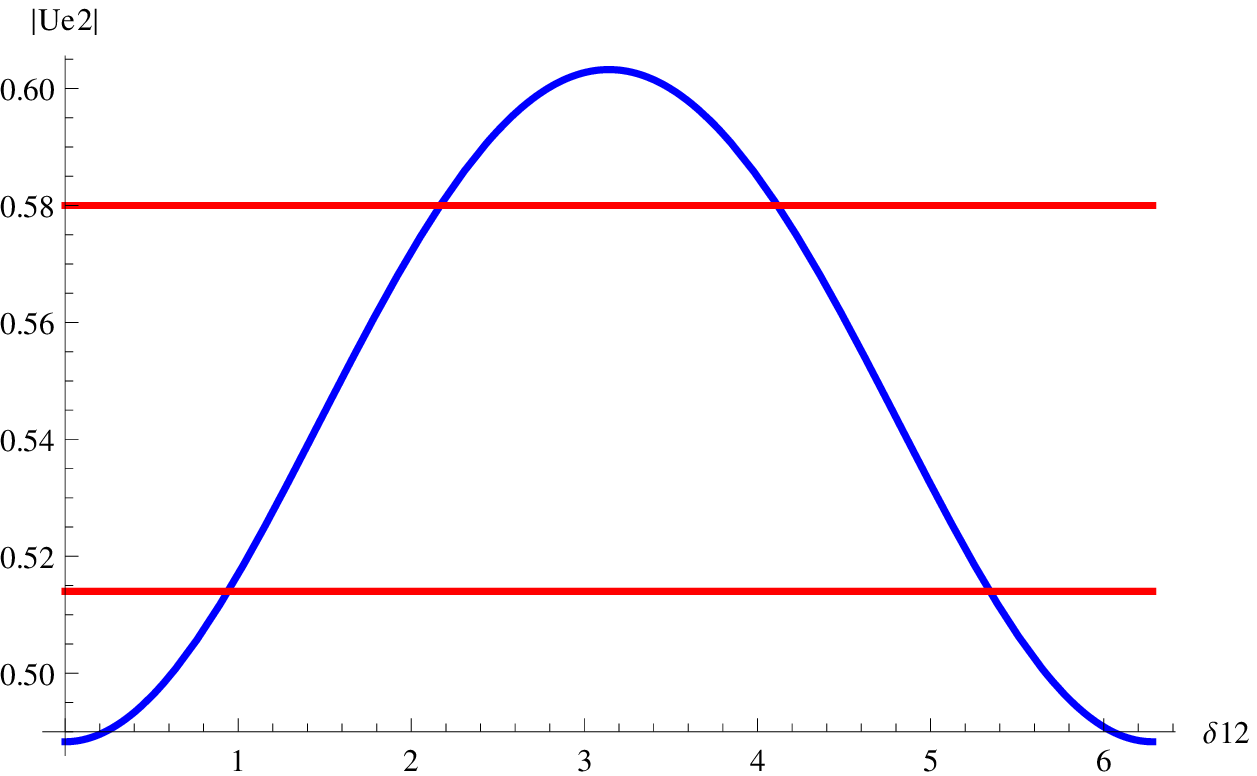}
\end{minipage}
\caption{Left: Evaluation of the three neutrino masses with the lightest
  mass (\(m_0\), in \(\eV\)). In the regime \(m_0<0.1\,\eV\) the
  assumption of a hierarchical pattern is indeed viable. Note also, that
  the ratio \(m_{\nu2}/m_{\nu3}\) basically does not change with
  decreasing \(m_0 = m_{\nu1}\). Right: The value of \(|U_{e2}|\) in
  dependency from \(\delta^\nu_{12}\)---the experimentally allowed
  \(3\,\sigma\) region (indicated by the horizontal red lines) is
  compatible with the choice \(\delta_{12}^\nu = \frac{\pi}{2}\), while
  not with \(\delta_{12}^\nu= 0\) or \(\pi\).}
\end{figure}

\paragraph{$U_{\text{PMNS}}^{\text{th}}$ as implied by the four leptonic
  mass ratios}
Albeit the hierarchy is not as strong as for quarks and charged
neutrinos, we dare to use the same description and show that indeed
large mass ratios in the four mass ratio parametrization also lead to
large mixing angles. The applicability of the whole method depends on
hierarchical masses. In~\ref{app:applic} we give a simple criterion
parameter to check whether the lower-rank approximations are good
approximations. Indeed, the deviation from unity is only a few
percent. Therefore, we safely use the previous described procedure.

With the predicted neutrino masses (which only know about $|U_{e2}|$)
and the knowledge of the charged fermion mass ratios, the leptonic
mixing matrix exhibits the following numerical values
\begin{eqnarray}\label{eq:prePMNS}
  |U_{\text{PMNS}}^{\text{th}}| =
  \begin{pmatrix}
    0.83^{+0.04}_{-0.05} & 0.54^{+0.06}_{-0.09} & 0.14 \pm 0.03 \\
    0.38^{+0.04}_{-0.06} & 0.57^{+0.03}_{-0.04} & 0.73 \pm 0.02 \\
    0.41^{+0.04}_{-0.06} & 0.61^{+0.03}_{-0.04} & 0.67 \pm 0.02
  \end{pmatrix},
\end{eqnarray}
whereas the implied amount of CP violation is displayed as
\begin{eqnarray}
  J_\ell = \Im(U_{e2}U_{\mu 3}U_{e 3}^*U_{\mu 2}^*) = 0.031^{+0.006}_{-0.007} .
\end{eqnarray}

We remark an astonishingly good agreement with the measured values (see
\ref{app:data}) and observe a close-to-maximal CP violation in the
lepton sector! (\(\delta_\text{CP} = 70^\circ\) from the central values:
\(J_\ell = J^\text{max}_\ell \sin\delta_\text{CP}\), the error on
\(J^\text{max}_\ell\) is nevertheless compatible with maximal CP
violation, \(\delta_\text{CP} = 90^\circ\).)

\subsection{About precision}
The goal of the presented work is not to be a precision analysis of
quark and lepton mixing. The projected values of the mixing matrices are
rather a rough-and-ready estimate compatible though very well with
experimental data. We wanted to show that the knowledge of fermion
masses is sufficient to describe their mixing accepting a hierarchical
nature.

The errors that are presented in Eqs.~\eqref{eq:postCKM} and
\eqref{eq:prePMNS} follow from the uncertainties in the masses. Better
precision in the determination of quark masses leads to better
discrimination in future whether the described procedure is valid. The
estimates are not too bad, nevertheless, we ignored radiative
corrections to the mixing matrices and constrain ourselves on a
tree-level discussion. One-loop corrections to the masses or Yukawa
couplings would be suppressed by factors \(Y_{ij} Y_{jk} Y_{kl} /
(16\pi^2)\) and are therefore in the range of the errors for the
masses. Renormalization group running of the parameters is also
negligible: quark mixing angles do basically not run. The running of
fermion mixing parameters depends on a factor \((m_{i} + m_{j} ) /
(m_{i} - m_{j})\) which is small for the hierarchical
spectra. Especially neutrino masses and mixings run only slightly in the
scenario which is under consideration in this work.

\section{Conclusions}
We investigated the long-standing question of understanding the
functional description of the mixing matrices in terms of the fermion
masses. The pure phenomenological observation of strong hierarchies
among the charged fermion masses $m_{f,3} \gg m_{f,2} \gg m_{f,1}$
guides the way to a parametrization of fermion mixings in terms of mass
ratios without further assumptions.  By solely exploiting the
mathematical properties of the mass matrices, namely their Singular
Value Decomposition, and making use of the double mass hierarchy pattern
(DMHP), we have shown that four mass ratios in each fermion sector and a
maximal CP violating phase in the 1-2 rotation are sufficient to
reproduce the numerical quantities of the fermionic mixing
matrices. Hierarchical masses guarantee a unique decomposition into
singular vectors up to a complex phase shared by the respective pair of
singular vectors of a singular value. This uniqueness theorem dissolves
the common ambiguities found in the literature originated in the freedom
of weak bases. Schmidt-Mirsky's approximation theorem has been used to
approximate the hierarchical mass matrices by lower-rank matrices that
are the closest one to the given full rank matrix. The connection of
each lower rank approximation to the nature of the Yukawa interactions,
$m_{f,i} = 0 \rightarrow Y^f_{ij} = 0 = Y_{ji}^f$, helps to simplify the
reparametrization of the mass matrix without loosing track of the
parameters. This connection is established via the minimal breaking of
maximal flavor symmetry \(\left[\U(3)\right]^3 \to \left[\U(2)\right]^3
\to \left[\U(1)\right]^3 \rightarrow \U(1)_F\) in each fermion sector,
where the remnant \(\U(1)_F\) symmetry is either baryon or lepton
number. The approximation, however, neglects sizeable terms in the mass
matrices that have been consistently added by use of correcting
rotations. The arbitrariness of complex phases is reduced by requiring
them to be either maximally CP violating (\(\pi/2\)) or CP conserving
(\(\pi\)). This assumption is motivated by the fact that the four mass
ratios should be enough to serve as mixing parameters in the unitary
\(3\times 3\) mixing matrix.

We found a remarkably good agreement of the projected magnitudes of both
the CKM and PMNS matrix elements and reproduce the Jarlskog invariant of
the quark sector quite well. The strength of this description in terms
of mass ratios lies in its invertibility. In the leptonic sector, we
have calculated the neutrino mass spectrum following from the inversion
of the formulae in the 1-2 mixing sector and the measured mass squared
differences. The lightest neutrino has a mass well below \(0.01\,\eV\),
while the largest neutrino mass lies around \(0.05\,\eV\). We therefore
conclude that, if also in the neutrino sector the mixing is determined
by the mass ratios without any further contribution, the electron
neutrino mass escapes its nearby measurement from tritium
decay. Moreover, we give a prediction for the leptonic CP phase close to
maximal, \(\delta^\nu_\text{CP} \approx 90^\circ\).

Hence, contrary to the common expectation, leptonic mixing angles are
found to be determined solely by the four leptonic mass ratios:
$m_e/m_\mu$, $m_\mu/m_\tau$, $m_{\nu 1}/m_{\nu 2}$, and $m_{\nu
  2}/m_{\nu 3}$ without any relation to the geometrical factors observed
in most flavor models.  Notwithstanding, we see a great power of the
described method in the application to flavor model building: once a
model gives hierarchical masses, the mixing follows from this
hierarchy. In contrast, our approach gives viable patterns and textures
for mass matrices in terms of the singular values (fermion masses). We
explicitly leave the question of a model behind open. Likewise, the
origin of CP violation stays unexplained, though our observation about
the distribution of CP phases gives an important starting point.

\section*{Acknowledgements}
UJSS would like to acknowledge useful and detailed discussions about
this idea in its initial stage with A.~Mondrag\'on and E.~Jim\'enez. The
authors want to thank the following people for their critical assessment
on this work during its different stages: M.~Spinrath, U.~Nierste,
C.~Wiegand, M.~Zoller, J.~Hoff, M.~Hoeschele, and K.~Melnikov. Also, the
authors are indebted to U.~Nierste for a careful reading of the
manuscript and his detailed comments on it. UJSS wants to acknowledge
financial support from the Karlsruhe House of Young Scientists (KHYS)
for his stay in Karlsruhe. WGH acknowledges support by the DFG-funded
research training group GRK~1694 ``Elementarteilchenphysik bei
h\"ochster Energie und h\"ochster Pr\"azision''. For last, UJSS is
grateful to the Institut f\"ur Theoretische Teilchenphysik (TTP) for its
warm hospitality during the realization and completion of this work.

\appendix

\section{State of the art in the fermion masses and mixing matrices}
\label{app:data}
In this section, we collect the current knowledge about fermion mixing
data and specify the input values we use in the following for the
masses.

For all numerical evaluations made in this work, we stick to the updated
values of the quark mixing matrix~\cite{Agashe:2014kda},
\begin{equation}\label{eq:CKMPDG}
  |V_{\text{CKM}}| =
  \begin{pmatrix}
    0.97427 \pm 0.00014 & 0.22536 \pm 0.00061 & 0.00355\pm 0.00015 \\
    0.22522 \pm 0.00061 & 0.97343 \pm 0.00015 & 0.0414 \pm 0.0012 \\
    0.00886^{+0.00033}_{-0.00032} & 0.0405^{+0.0011}_{-0.0012} &
    0.99914\pm 0.00005 \end{pmatrix},
\end{equation}
with the Jarlskog invariant equal to $ J_q =
(3.06^{+0.21}_{-0.20})\times 10^{-5}$. In the standard parametrization
by the Particle Data Group (PDG), the central values give the following
mixing angles,
\begin{equation}
  \theta_{12}^q \approx 13.3^\circ, \qquad
  \theta_{13}^q \approx 0.2^\circ, \qquad
  \theta_{23}^q \approx 2.4^\circ.
\end{equation}

The most recent update on the $3\sigma$ allowed ranges of the elements
of the PMNS mixing matrix are given by~\cite{Gonzalez-Garcia:2014bfa},
\begin{equation}
  |U_{PMNS}| =
  \begin{pmatrix}
    0.801\rightarrow 0.845 & 0.514 \rightarrow 0.580 & 0.137 \rightarrow 0.158 \\
    0.225 \rightarrow 0.517 & 0.441 \rightarrow 0.699 & 0.614 \rightarrow 0.793 \\
    0.246 \rightarrow 0.529 & 0.464 \rightarrow 0.713 & 0.590
    \rightarrow 0.776
  \end{pmatrix}.
\end{equation}
Where the best fit points of the mixing angles are
\begin{equation}
  \theta_{12}^\ell = 33.48^\circ, \qquad
  \theta_{13}^\ell = 8.50^\circ, \qquad
  \theta_{23}^\ell = 42.3^\circ.
\end{equation}
The maximal value of the leptonic Jarlskog invariant is given by
\(J_\ell^\text{max} = 0.033 \pm 0.010\) and different from zero at more
than \(3\,\sigma\)---still, the proper \(J_\ell\) has first to be
multiplied by \(\sin\delta_\mathrm{CP}\) and is supposed to be smaller.

The study of the mixing matrices in terms of the masses is done at the
scale of the $Z$ boson mass.  The input values for the numerical
calculations are obtained using the experimental values of the quark
masses as given by the PDG Review 2014~\cite{Agashe:2014kda} and running
them to the scale of the \(Z\) boson determining the electroweak
scale. We include highest precision running in QCD by the virtue of the
RunDec package \cite{Chetyrkin:2000yt}. For completeness, we show the
input values and their uncertainties as well as the resulting outputs in
Table~\ref{Table:Quark-masses}.

\begin{table}[tb]
    \caption{The quark masses are run to the $Z$ boson mass scale by
      virtue of the RunDec package~\cite{Chetyrkin:2000yt}. The mass
      inputs correspond to the experimental measured values while the
      outputs, evaluated at the \(Z\) pole, include the resummation of
      higher order corrections from QCD by the RG running. RunDec takes
      properly into account the decoupling of heavy quarks below their
      scale. All masses are given in \(\GeV\).}
    \label{Table:Quark-masses}
  \begin{center}
    \ra{1.3}
    \begin{tabular}{cc}
      \toprule
      \textbf{input} &\textbf{output} \\
      \midrule
      \(m_u(2\,\GeV) = 0.0023^{+0.0007}_{-0.0005}\) & \(m_u(M_Z) =
      0.0013^{+0.0004}_{-0.0003}\) \\
      \(m_d(2\,\GeV) = 0.0048^{+0.0005}_{-0.0003}\) & \(m_d(M_Z) =
      0.0028^{+0.0003}_{-0.0002}\) \\
      \(m_s(2\,\GeV) = 0.095 \pm 0.005\) & \(m_s(M_Z) =
      0.055 \pm 0.003\) \\
      \(m_c(m_c) = 1.275 \pm 0.025\) & \(m_c(M_Z) =
      0.622 \pm 0.012\) \\
      \(m_b(m_b) = 4.18 \pm 0.03\) & \(m_b(M_Z) =
      2.85 \pm 0.02\) \\
      \(m_t(\mathrm{OS}) = 173.07 \pm 1.24\) & \(m_t(M_Z) =
      172.16^{+1.47}_{-1.46}\) \\
      \bottomrule
    \end{tabular}
  \end{center}
\end{table}

The reported measured on-shell values in MeV for the charged lepton masses are,
\begin{eqnarray}
  m_e=0.510998928, \quad\quad m_\mu=105.6583715, \quad\quad m_\tau=1776.82 \pm 0.16,
\end{eqnarray}
where we have neglected the tiny experimental errors in the first two
generation masses. The recent changes of this values affect only the
few last digits. Therefore, we safely trust the results of
\cite{Xing:2007fb} for their values at the \(Z\) scale (in \(\MeV\)):
\begin{eqnarray}\label{eq:leptmass}
  m_e(M_Z)=0.486570161, \quad\quad
  m_\mu(M_Z)=102.7181359, \quad\quad
  m_\tau(M_Z)=1746.24^{+0.20}_{-0.19}.
\end{eqnarray}

The nine mass ratios are of essential use in the evaluation of the
analytic formulae to describe fermion mixing. We show our input values
determined from Table~\ref{Table:Quark-masses} and Eq.~\eqref{eq:leptmass} in
Table~\ref{Table:Q-massratios}.

\begin{table}[tb]
    \caption{Charged fermions mass ratios at the $M_Z$
      scale.}
    \label{Table:Q-massratios}
  \begin{center}
    \ra{1.3}
    \begin{tabular}{cccc}
      \toprule
      $f$ & $m_{f,1}/m_{f,2}$ & $m_{f,1}/m_{f,3}$ & $m_{f,2}/m_{f,3}$ \\
      \midrule
      $u$ & $0.0021^{+0.0007}_{-0.0005}$ &
      $(7.6^{+2.4}_{-1.8} )\times 10^{-6}$ & $0.0036 \pm 0.0001$ \\
      $d$ & $0.051^{+0.009}_{-0.006}$ &
      $(9.8^{+1.1}_{-0.7})\times 10^{-4}$ & $0.019\pm 0.0012$ \\
      $e$ & 0.00474 & 0.000279 & 0.0588 \\
      \bottomrule
    \end{tabular}
  \end{center}
\end{table}

In the case of neutrinos, only two squared mass differences have been
measured whose values are taken from~\cite{Gonzalez-Garcia:2014bfa},
\begin{equation}\label{eq:Deltam2}
  \begin{aligned}
    {\text{NO:}} \quad \Delta m_{31}^2 = +2.457 \pm 0.002 \times 10^{-3}\,\eV^2,
    \quad\quad&
    {\text{IO:}} \quad \Delta m_{32}^2 = -2.448 \pm 0.047 \times 10^{-3}\,\eV^2,\\
    \Delta m_{21}^2 = 7.50^{+0.19}_{-0.17} &\times 10^{-5}\,\eV^2,
  \end{aligned}
\end{equation}
where NO and IO stand for normal and inverted ordering,
respectively.

Still, the most recent direct bound on the neutrino mass scale stems
from tritium beta decay experiments: \(m(\nu_e) \lesssim 2\,\eV\) at
$95\%$ C.L.~\cite{Aseev:2011dq}. The KATRIN experiment is going to
improve this bound by one order of magnitude~\cite{Osipowicz:2001sq}.

\section{Applicability of the method}
\label{app:applic}
The Schmidt-Mirsky theorem relates the validity of the lower rank
approximation to a measure of being close to the full rank matrix. This
measure has to be a scalar parameter and can be any norm. In the
original formulation, the Frobenius norm was used, which is also the
most natural choice since it is the square root over the sum of squared
singular values and directly related to one of the invariants of the
mass matrix
\begin{eqnarray}
  \parallel {\cal M}_f \parallel_{{\bf{F}}} \,\, = \sqrt{\sum_{i=1,2,3} m_{f,i}^2}.
\end{eqnarray}
The use of this norm serves as a way to define a criterion which allows
us to distinguish when the hierarchy is strong enough as to safely make
an approximation. In this regard, we define the parameter $x_f^r$ as,
\begin{eqnarray} \label{eq:crit-parametr} x_f^r \equiv
  \frac{\sqrt{(r-1)m_{f,2}^2 + m_{f,3}^2}}{\parallel {\cal
      M}_f \parallel_{{\bf{F}}}} = \sqrt{\frac{(r-1)m_{f,2}^2 +
      m_{f,3}^2}{m_{f,1}^2+m_{f,2}^2+m_{f,3}^2}},
\end{eqnarray}
where $r={\operatorname{rank}}[{\cal M}^r_f] \in \lbrace 1,
2\rbrace$. The approximation becomes better the closer \(x^r_f\) is to
one and is exact in the $x_f^r \rightarrow 1$
limit. Eq.~\eqref{eq:crit-parametr} is actually the ratio of the lower
rank approximated mass matrix norm with the original norm.  Hence,
$x_f^r$ is a measure of the applicability of the
method. Table~\ref{Table:xrf} shows the different values obtained of
$x_f^r$ for the several charged fermion masses. The values in the rank
one approximation, $r=1$, for all practical purposes equal to one,
though for both charged and neutral leptons deviate in the per mill and
percent regime, respectively. From here we can already understand why the
quark mixing matrix is so close to the unit matrix which is the trivial
mixing matrix in the rank one approximation. In a similar manner, the
very mild hierarchy for neutrinos leads to a stronger deviation from the
rank one approximation and therefore larger mixing angles.

\begin{table}[tb]
  \caption{Values of the criterion parameter $x_f^r \equiv
    \sqrt{[(r-1)m_{f,2}^2 + m_{f,3}^2
      ]/(m_{f,1}^2+m_{f,2}^2+m_{f,3}^2)}$, for the different cases of
    the fermion masses, where $x_f^r$ provides a measure of
    the applicability of the method. The fact that all cases here are
    sufficiently close to one guarantees the safe use of the lowest
    rank approximations. Even for neutrinos, \(x^2_\nu\) is close to
    one, where we exploit the prediciton for neutrino masses from Sec.~\ref{sec:lept}.}
  \label{Table:xrf}
  \begin{center}
    \begin{tabular}{ccccc} \toprule $x^{r}_f$& $u$ & $d$ & $e$ &\(\nu\)\\
      \midrule
      $r=1$ & 0.999993 & 0.999816 & 0.998274 & 0.978894\\
      $r=2$ & 0.999999 & 0.999999 & 0.999999 & 0.996773\\
      \bottomrule
    \end{tabular}
  \end{center}
\end{table}

\section{Hierarchical mass matrices}
\label{app:m11}
We show how to derive the hierarchical structure of the mass
  matrices by the use of the lower-rank approximation theorem and the
  principle of minimal flavor violation. Let us consider the two-flavor
  case and the mass matrix
\begin{eqnarray}
  \bf m = \begin{pmatrix}
    m_{ss} & m_{sl} \\ m_{ls} & m_{ll}
  \end{pmatrix},
\end{eqnarray}
with the two singular values $\sigma_s$ and $\sigma_l$ respecting the
hierarchy \(\sigma_s \ll \sigma_l\).

We decompose the mass matrix in terms of the Singular Value decomposition
\begin{eqnarray} \label{eq:svd}
  \mathbf{L}\,\mathbf{m}\,\mathbf{R}^\dagger = \diag(\sigma_s, \sigma_l),
\end{eqnarray}
where the left and right unitary matrices diagonalize the Hermitian
products
\begin{eqnarray}\label{eq:LRdiag}
  \mathbf{L}\, \mathbf{m}\,\mathbf{m}^\dagger\, \mathbf{L}^\dagger =
  \begin{pmatrix}
    \sigma^2_s & 0 \\ 0 & \sigma^2_l
  \end{pmatrix}
  = \mathbf{R}\, \mathbf{m}^\dagger\, \mathbf{m}\, \mathbf{R}^\dagger.
\end{eqnarray}
Each Hermitian product can be expressed as a sum of rank one matrices
with the components of $\mathbf{L}$ and $\mathbf{R}$,
\begin{eqnarray}\label{eq:left}
  \mathbf{m}\,\mathbf{m}^\dagger  = \sigma_s^2
  \begin{pmatrix}
    |L_{11}|^2 & L_{11}L_{21}^* \\ L^*_{11}L_{21} & |L_{21}|^2
  \end{pmatrix} +
  \sigma_l^2
  \begin{pmatrix}
    |L_{12}|^2 & L_{12}L_{22}^* \\ L^*_{12}L_{22} & |L_{22}|^2
  \end{pmatrix}
\end{eqnarray}
and
\begin{eqnarray}
  \mathbf{m}^\dagger\, \mathbf{m}  = \sigma_s^2
  \begin{pmatrix}
    |R_{11}|^2 & R_{11}R_{21}^* \\ R^*_{11}R_{21} & |R_{21}|^2
  \end{pmatrix} +
  \sigma_l^2
  \begin{pmatrix}
    |R_{12}|^2 & R_{12}R_{22}^* \\ R^*_{12}R_{22} & |R_{22}|^2
  \end{pmatrix}.
\end{eqnarray}
Due to our lack of knowledge of right-handed flavor mixing, the relevant
object that determines our phenomenology is the left Hermitian product,
\(\mathbf{m}\, \mathbf{m}^\dagger\).

\paragraph{Applying Schmidt-Mirsky's approximation theorem}
Consider the rank-one approximation in Eq.~\eqref{eq:left} by
\(\hat\sigma = \sigma_s / \sigma_l = 0\) normalized with respect to the
larger singular value
\begin{eqnarray}
  \hat{\mathbf{m}}^{r=1} (\hat{\mathbf{m}}^{r=1})^\dagger  =
  \begin{pmatrix}
    |L_{12}|^2 & L_{12}L_{22}^* \\ L^*_{12}L_{22} & |L_{22}|^2
  \end{pmatrix}.
\end{eqnarray}
The components of the left unitary matrix depend on \(\hat\sigma\). In
the limit \(\hat\sigma \to 0\), there is trivial mixing and the rank one
left Hermitian product is
\begin{eqnarray}
  \hat{\mathbf{m}}^{r=1}(\hat{\mathbf{m}}^{r=1})^\dagger  =
  \begin{pmatrix}
    0 & 0 \\ 0 & 1
  \end{pmatrix}.
\end{eqnarray}

A small breaking of the \([\U(1)]^2\) symmetry for the massless fermions
implies only a small deviation from the trivial mixing:
\begin{eqnarray}
  |\mathbf{L}| \sim \begin{pmatrix}
    1 & \theta \\  \theta & 1
  \end{pmatrix}.
\end{eqnarray}
The mixing angle is related to the parameter of symmetry breaking
\(\hat\sigma\) and it is an easy exercise to derive \(\theta \sim
\sqrt{\hat\sigma}\) from Eq.~\eqref{eq:LRdiag}.

We then get an estimate on the magnitudes of each element in Eq.~\eqref{eq:left}
\begin{eqnarray} \label{apC-mmdagger}
  |\hat{\mathbf{m}}\,
  \hat{\mathbf{m}}^\dagger| \sim
  \begin{pmatrix}
    \mathcal{O}(\theta^2) & \mathcal{O}(\theta) \\
    \mathcal{O}(\theta) & 1 + \mathcal{O}(\theta^2)
  \end{pmatrix}.
\end{eqnarray}

The explicit form of the mass matrix \(\mathbf{m}\) stays unknown as
long as we have no information about \(\mathbf{R}\). However, the
minimal breaking of the maximal flavor symmetry applies to both
chiralities simultaneously and the argument from above is the same for
the right Hermitian product. We therefore know that \(\mathbf{L}\) and
\(\mathbf{R}\) have the same moduli and get the hierarchical structure
of \(\mathbf{m}\):
\begin{eqnarray}
  \hat{\mathbf{m}} =
  \begin{pmatrix}
    m_{ss} & m_{sl} \\
    m_{ls} & m_{ll}
  \end{pmatrix} \sim \begin{pmatrix}
    {\cal O}(\theta^2) & {\cal O}(\theta) \\
    {\cal O}(\theta) & 1 + \mathcal{O}(\theta^2)
  \end{pmatrix},
\end{eqnarray}
with $|m_{sl}| = |m_{ls}|$ as a natural consequence of hierarchical
masses and minimal flavor symmetry breaking. The hierarchical structure
for the mass matrix and its Hermitian product is the same. Hence, due to
the strong hierarchy in the masses we can neglect the role of
\(|m_{ss}|^2 \sim \theta^4\) in~\eqref{apC-mmdagger} working with the
leading order contributions in \(\theta\) and assume \(m_{ss} = 0\) as
done in Eq.~\eqref{eq:approxmassmat}. This gives corrections to the
Gatto-Satori-Tonin relation, \(\tan\theta = \sqrt{\sigma_s / \sigma_l} =
\sqrt{\hat\sigma}\), which are \(\mathcal{O}(\theta^3) =
\mathcal{O}(\hat\sigma\sqrt{\hat\sigma})\) and therefore neglected.

\section{Explicit approximate formulae for the mixing angles and the
  Jarlskog invariant}
\label{app:Formulae}

The explicit formulae for the distinct mixing matrix elements in terms
of the mass ratios is rather lengthy.  We opt then, to show only the
three mixing angles, used in the standard parametrization, with the
corresponding Jarlskog invariant. This allows to express the mixing
angles in terms of three moduli of the mixing matrix
\begin{eqnarray}
  \sin\theta_{23}^{f=q,\ell} =
  \frac{\left|V^{f=q,\ell}_{23}\right|}{\sqrt{1-\left|V_{13}^{f=q,\ell}\right|^2}},
  \quad
  \sin\theta_{12}^{f=q,\ell} =
  \frac{\left|V^{f=q,\ell}_{12}\right|}{\sqrt{1-\left|V_{13}^{f=q,\ell}\right|^2}},
  \quad
  \sin\theta_{13}^{f=q,\ell} = \left|V^{f=q,\ell}_{13}\right|\;.
\end{eqnarray}
In the four mass ratios parametrization it is more natural to give not the formulae
of the mixing angles in terms of the masses but rather of the aforementioned moduli
\begin{align}
|V_{12}^{f=q,\ell}| &\approx \sqrt{\frac{\hat{m}^a_{12}+\hat{m}^b_{12}}{(1+\hat{m}^a_{12})(1+\hat{m}^b_{12})}}
\label{eq:theta12}\;, \\
  |V^{f=q,\ell}_{23}| &\approx \mp
  \frac{\sqrt{\hat{m}_{13}^a}+\sqrt{\hat{m}_{13}^b}+\sqrt{\hat{m}_{23}^a}\mp
    \sqrt{\hat{m}_{23}^b}+\sqrt{\hat{m}_{13}^a\hat{m}_{23}^a}\pm
    \sqrt{\hat{m}_{13}^b\hat{m}_{23}^b}}
  {\sqrt{(1+\hat{m}_{13}^a)(1+\hat{m}_{13}^b)(1+\hat{m}_{23}^a)(1+\hat{m}_{23}^b)(1+\hat{m}_{13}^a\hat{m}_{23}^a)
      (1+\hat{m}_{13}^b\hat{m}_{23}^b)}}
  \label{eq:theta23}\;, \\
  |V_{13}^{f=q,\ell}| &\approx
  \mp|V_{23}^{f=q,\ell}|\sqrt{\frac{\hat{m}_{12}^a}{1+\hat{m}_{12}^a}} \;+
\nonumber \\ &\;
\frac{\sqrt{\hat{m}^a_{13}}-\sqrt{\hat{m}^b_{13}}+\sqrt{\hat{m}^a_{13}\hat{m}^a_{23}}+
  \sqrt{\hat{m}^b_{13}\hat{m}^b_{23}}+ \hat{m}^a_{23}+\hat{m}^b_{23}}
{\sqrt{(1+\hat{m}_{13}^a)(1+\hat{m}_{13}^a\hat{m}_{23}^a)\left(1+(\hat{m}_{23}^a)^2\right)(1+\hat{m}_{12}^a)(1+\hat{m}^b_{13})(1+\hat{m}^b_{13}\hat{m}^b_{23})\left(1+(\hat{m}^b_{23})^2\right)}}\;,
\label{eq:theta13}
\end{align}
where we have denoted $\hat{m}^{a(b)}_{ij} = m^{a(b)}_i / m^{a(b)}_j$,
the upper and lower signs in Eq.~\ref{eq:theta23} correspond to $q$ and
$\ell$, respectively. The two fermion species of each sector are $a=
u,e$ and $b = d,\nu$.

The Jarlskog invariant is given by,
\begin{eqnarray}
  J_{f=q,\ell} \approx \cos\theta_{12}^b \sin\theta_{12}^b
  \sin\theta_{23}^{f=q,\ell}
  \left(\sin\theta_{12}^a\sin\theta_{23}^{f=q,\ell}
    + \sin\theta_{13}^a-\sin\theta_{13}^b\right),
\end{eqnarray}
where
\begin{eqnarray}
  \sin\theta_{12}^{a(b)} =
  \sqrt{\frac{\hat{m}^{a(b)}_{12}}{1+\hat{m}^{a(b)}_{12}}}
  \quad {\text{and}} \quad
  \sin\theta_{13}^{a(b)} \approx \frac{\pm\sqrt{\hat{m}^{a(b)}_{13}}+ \sqrt{\hat{m}^{a(b)}_{13}\hat{m}^{a(b)}_{23}} + \hat{m}^{a(b)}_{23}}
  {\sqrt{\left(1+\hat{m}^{a(b)}_{13}\right)\left(1+\hat{m}^{a(b)}_{13}\hat{m}^{a(b)}_{23}\right)\left(1+(\hat{m}^{a(b)}_{23})^2\right)}}.
\end{eqnarray}
The approximate relations here given differ from the complete one in
$\sim 1$\% order.

\bibliography{dmhp}
\bibliographystyle{utcaps}

\end{document}